\documentclass[sn-basic]{sn-jnl}

\usepackage{graphicx}%
\usepackage{multirow}%
\usepackage{amsmath,amssymb,amsfonts}%
\usepackage{amsthm}%
\usepackage{mathrsfs}%
\usepackage[title]{appendix}%
\usepackage{xcolor}%
\usepackage{textcomp}%
\usepackage{manyfoot}%
\usepackage{booktabs}%
\usepackage{algorithm}%
\usepackage{algorithmicx}%
\usepackage{algpseudocode}%
\usepackage{listings}%

\theoremstyle{thmstyleone}%
\theoremstyle{thmstyletwo}%

\theoremstyle{thmstylethree}%
\newtheorem{definition}{Definition}%

\begin{document}

\title[HiMARS: Hybrid multi-objective algorithms for recommender systems]{HiMARS: Hybrid multi-objective algorithms for recommender systems}


\author[1]{\fnm{Elaheh} \sur{Lotfian}}\email{e.lotfian@gmail.com}
\author[2]{\fnm{Alireza} \sur{Kabgani}}\email{alireza.kabgani@uantwerp.be}

\affil[1]{ \orgaddress{\city{Antwerp}, \country{Belgium}}}

\affil[2]{\orgdiv{Department of Applied Mathematics}, \orgname{University of Antwerp}, \orgaddress{\city{Antwerp}, \country{Belgium}}}


\abstract{In recommender systems, it is well-established that both accuracy and diversity are crucial for generating high-quality recommendation lists. However, achieving a balance between these two typically conflicting objectives remains a significant challenge. In this work, we address this challenge by proposing four novel hybrid multi-objective algorithms inspired by the Non-dominated Neighbor Immune Algorithm (NNIA), Archived Multi-Objective Simulated Annealing (AMOSA), and Non-dominated Sorting Genetic Algorithm-II (NSGA-II), aimed at simultaneously enhancing both accuracy and diversity through multi-objective optimization.
Our approach follows a three-stage process: First, we generate an initial top-$k$ list using item-based collaborative filtering for a given user. Second, we solve a bi-objective optimization problem to identify Pareto-optimal top-$s$ recommendation lists, where $s \ll k$, using the proposed hybrid algorithms. Finally, we select an optimal personalized top-$s$ list from the Pareto-optimal solutions.
We evaluate the performance of the proposed algorithms on real-world datasets and compare them with existing methods using conventional metrics in recommender systems such as accuracy, diversity, and novelty. Additionally, we assess the quality of the Pareto frontiers using metrics including the spacing metric, mean ideal distance, diversification metric, and spread of non-dominated solutions. Results demonstrate that some of our proposed algorithms significantly improve both accuracy and diversity, offering a novel contribution to multi-objective optimization in recommender systems.}




\maketitle

\section{Introduction}
\label{sec:intro}
The rapid growth of digital content and data-driven services has led to an era where users are often overwhelmed by the sheer volume of available information. 
From online marketplaces to streaming platforms, navigating through this vast array of content has made the development of effective filtering and recommendation tools indispensable. 
Among these tools, recommender systems have emerged as a key solution for managing items presented by online services, such as movies, news, and products. These systems help users find relevant content by generating personalized suggestions, addressing the challenge of information overload and enhancing overall user satisfaction \citep{Bobadilla13,Lu12}.

Recommender systems have applications across various domains, including 
e-commerce \citep{Jiang2019}, 
large language models \citep{Zhao24}, 
social networks \citep{Zhao2018}, 
and news platforms  \citep{Raza2022}, among others \citep{Ko2022}. 
As these systems become integral to many digital services, research has primarily focused on improving the accuracy of recommendations \citep{Jiang2019,Wu23}.  However, focusing solely on accuracy may introduce significant drawbacks: these systems often suggest highly similar items, limiting practical utility \citep{Zheng21,Zheng22}. This issue can appear as the long-tail problem, where items rated by only a few users rarely appear in future recommendations \citep{Park08}. Similarly, in news recommendations, focusing solely on accuracy can lead to filter bubbles, presenting users with increasingly homogeneous content and isolating them within their own ideological or cultural views \citep{Michiels23}.

To address these issues, it is crucial to consider factors beyond accuracy, such as diversity \citep{Kunaver17}. Incorporating diversity ensures that users are exposed to a broader range of content, improving the system's overall effectiveness. However, achieving a balance between accuracy and diversity is inherently challenging, as improving one often diminishes the other, making it difficult to optimize both objectives simultaneously \citep{Lacerda17}.

Given this inherent trade-off, reframing recommender system optimization as a multi-objective problem has become essential. Multi-objective optimization methods allow simultaneous consideration of conflicting objectives, such as accuracy and diversity, and have been extensively studied in recommender systems \citep{Cai20,Chai21,Cui17,Hamedani19,Lacerda17,Ma23,Zaizi23,Zheng22,Zou21}. Among the commonly used approaches are scalarization techniques and Pareto dominance methods. Scalarization methods, though straightforward, are often ineffective for non-convex optimization problems \citep{Drugan13,Ehrgott}.

In contrast, Pareto-based approaches provide a set of Pareto solutions (see Definition~\ref{def:par}), offering multiple recommendation lists that balance accuracy and diversity. However, they introduce two significant challenges: (1) developing methods that generate high-quality Pareto-optimal lists, and (2) creating systematic approaches to select the most tailored recommendation list from a potentially large Pareto set \citep{Wang16}. Addressing these challenges is essential for the practical implementation of Pareto-based methods in recommender systems \citep{Chai21}.

Furthermore, evaluating multi-objective algorithms requires careful consideration of appropriate metrics. While accuracy, diversity, and novelty are the traditional metrics for assessing recommender systems \citep{Avazpour14,Geng15,Wang16}, these alone are insufficient for evaluating the quality of Pareto frontiers. Additional metrics, such as Spacing Metric (SM), Mean Ideal Distance (MID), Diversification Metric (DM), and Spread of Non-Dominated Solutions (SNS), are crucial for assessing the uniformity and quality of the Pareto frontier (discussed further in Subsection~\ref{subsec:evParfront}).

In this paper, we propose four hybrid multi-objective algorithms designed to improve the balance between accuracy and diversity in recommender systems:
\begin{itemize}
  \item \textbf{HANv1 and HANv2}: These algorithms combine two well-established multi-objective optimization methods: Archived Multi-Objective Simulated Annealing (AMOSA) \citep{Bandyopadhyay} and Non-dominated Sorting Genetic Algorithm-II (NSGA-II) \citep{Deb}. While NSGA-II is effective at exploring a wide search space, AMOSA excels at exploiting local search areas. Prior studies have shown that although NSGA-II generates a broader Pareto frontier, AMOSA can partially dominate this frontier \citep{Lotfian23,Saadatpour20}. Recently, \cite{Lotfian23} proposed a hybrid AMOSA\_NSGA-II (HAN) algorithm to address multi-objective optimization in spatial sampling. The main idea behind HAN is to apply mutation and crossover operations \citep{Deb} to the archive obtained by AMOSA during each iteration, allowing HAN to produce diverse and high-quality solutions across the search space.

  \item \textbf{HANIv1 and HANIv2}: These algorithms integrate the Non-dominated Neighbor Immune Algorithm (NNIA) \citep{Gong08} with AMOSA. NNIA enhances exploration by cloning isolated Pareto points, thereby improving diversity within the population. While NNIA has been previously applied in multi-objective recommender systems \citep{Chai21,Geng15}, this is the first work to combine NNIA and AMOSA (HANI) to address the accuracy-diversity trade-off in this context.
\end{itemize}

We evaluate these algorithms with metrics such as accuracy, diversity, and novelty, as well as metrics that specifically assess Pareto frontier quality, including Spacing Metric (SM), Mean Ideal Distance (MID), Diversification Metric (DM), and Spread of Non-Dominated Solutions (SNS). Finally, we propose a method for selecting the most suitable recommendation list from the Pareto set, based on its proximity to an ideal solution.

The remainder of this paper is organized as follows: Section \ref{sec:lit} reviews the related literature, while Section \ref{sec:formulation} presents the problem formulation, methodology, and proposed algorithms. Section \ref{sec:evaluation} introduces the datasets and evaluation metrics used to assess the recommendation lists and Pareto frontiers, along with experimental results. Finally, Section \ref{sec:conclusion} provides further discussion and concludes the paper.

\section{Literature review}
\label{sec:lit}
To address the inherent conflict between accuracy and diversity in recommender systems, various multi-objective algorithms have been proposed. 
\cite{Geng15} introduced a recommendation algorithm based on the NNIA, which combines collaborative filtering with a Multi-Objective Evolutionary Algorithm (MOEA) to improve recommendation diversity while preserving accuracy. 
Similarly, \cite{Chai19} proposed a multi-objective recommender system that integrates singular value decomposition with a multi-objective immune algorithm. In a subsequent study, \cite{Chai21} enhanced this approach by incorporating the PROMETHEE method \citep{Brans85} to evaluate Pareto set solutions, yielding a ranked list of top recommendations.


\cite{Zou15} addressed the accuracy-diversity conflict by employing NSGA-II, utilizing probabilistic spreading methods and recommendation coverage metrics to evaluate accuracy and diversity, respectively. \cite{Huang20} applied NSGA-II within an e-commerce recommendation system, while \cite{Almarimi19} implemented it for web service API recommendations.

\cite{Wang16} focused on recommending long-tail items using a multi-objective framework that employed decomposition-based MOEAs.
 Similarly, \cite{Hamedani19} addressed the challenges of long-tail item recommendations by utilizing AMOSA in a three-objective optimization context.
 
\cite{Ribeiro12} proposed a hybrid recommendation approach that optimizes accuracy, diversity, and novelty through a multi-objective methodology, employing the SPEA-2 algorithm to compute weights and aggregate scores for each item. This work was later extended by \cite{Ribeiro15}, who utilized Pareto efficiency to integrate multiple recommendation algorithms, optimizing various objectives simultaneously.
 
\cite{Wang14} introduced a multi-objective evolutionary algorithm based on decomposition to jointly optimize predicted scores and popularity of items. 
\cite{Zhou13} tackled the accuracy-diversity dilemma by proposing a heat conduction algorithm, which leverages a ground user and a free parameter to maximize accuracy. 
\cite{Mikeli13} developed a model-based recommender system incorporating a novel multi-criteria approach, while \cite{Zhang08} proposed an optimization technique that enhances diversity by optimizing two objective functions: preference similarity and item diversity.

In addition to these multi-objective optimization approaches, other perspectives on balancing accuracy and diversity have been explored in the literature. For example, context-aware recommender systems (CARS) utilize techniques such as pre-filtering, post-filtering, and contextual modeling \citep{Adomavicius15}, although these methods do not explicitly frame the problem as a multi-objective optimization task. 

\section{Problem formulation and method}
\label{sec:formulation}
In this section, we present the framework for the proposed method, including problem formulation and the introduction of four hybrid multi-objective algorithms designed for recommender systems.

\subsection{Framework of HiMARS}

The proposed multi-objective recommender system, HiMARS, operates through a three-stage process. First, a candidate generation stage produces an initial top-$k$ list of promising items for the target user. This stage serves to reduce the search space from the entire item catalog to a computationally tractable size. This candidate list can be generated using a variety of established techniques, including collaborative filtering, such as item-based CF \citep{Sarwar01}, matrix factorization models, such as SVD \citep{Koren09}, content-based models \citep{Pazzani07}, and heuristics designed to inject novelty, such as selecting popular items with low exposure.
Critically, these methods can also be combined into a hybrid ensemble to create an even richer and more diverse candidate pool.
In this context, $k$ is significantly larger than $s$, the final size of the top-$s$ recommendation list, i.e., $s \ll k$.

In the second stage, the proposed hybrid algorithms optimize conflicting objectives, i.e., accuracy and diversity, to obtain a Pareto set of top-$s$ recommendation lists. These lists are sublists derived from the top-$k$ candidate list generated in the first stage.

\begin{figure}[h]
\centering
\includegraphics[width=0.6\textwidth]{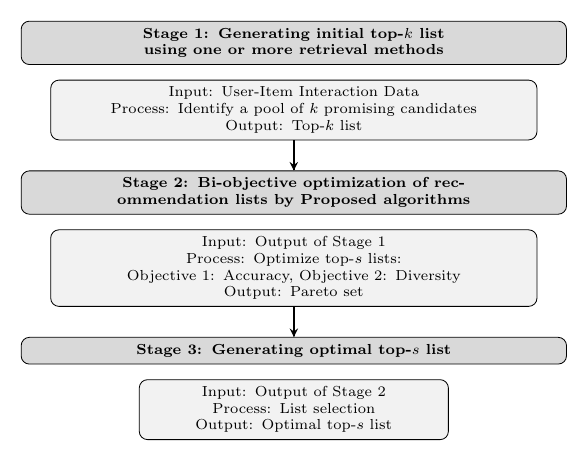}
\caption{Framework of HiMARS}\label{fig1draft2}
\end{figure}

Finally, in the third stage, we introduce an evaluation method to assess the recommendation lists within the Pareto set. This evaluation allows for the selection of an optimal top-$s$ list for each user, balancing the trade-offs between two objectives. This entire process is illustrated in Figure \ref{fig1draft2}.

\subsection{Problem formulation}
A general multi-objective optimization problem (MOP) can be formulated as follows:
\begin{align}
\label{eq3}
\max_{x\in \mathbb{R}^n} f(x)=(f_1 (x),\ldots,f_m (x)),
\end{align}
where $f_i:\mathbb{R}^n\rightarrow\mathbb{R}$ for each $i \in I_m:=\{1, \ldots, m\}$, with $m > 1$ objectives. In MOPs, the objectives typically conflict, meaning that an improvement in one objective often leads to deterioration in others. Since the objective space $\mathbb{R}^m$ is multi-dimensional, we cannot directly compare vectors using the standard ordering in $\mathbb{R}$. Instead, we use the notion of \textbf{Pareto dominance}. For two vectors $x, y \in \mathbb{R}^n$, we say that $x$ dominates $y$ and write it as $f(y) \lneqq f(x)$ if
\begin{itemize}
  \item $f(y_i) \leq f(x_i) \text{ for all } i \in \{1, \ldots, n\}$,
  \item $f(y_j) < f(x_j) \text{ for at least one } j$.
\end{itemize}
If $x$ and $y$ do not dominate each other, we say they are \textbf{non-dominated}, denoted as $x \nless\ngtr y$.

\begin{definition} \label{def:par}
\citep{Ehrgott}
A vector $\hat{x} \in \mathbb{R}^n$ is called a \textbf{Pareto point} of Problem (\ref{eq3}) if there is no $x \in \mathbb{R}^n$ such that $f(\hat{x}) \lvertneqq f(x)$. If $\hat{x}$ is a Pareto point, then $f(\hat{x})$ is called a \textbf{Pareto solution}. The set of all Pareto points is the \textbf{Pareto set}, and the corresponding image in the objective space is the \textbf{Pareto frontier}. The \textbf{ideal point} $I_P$ with respect to the Pareto frontier $P$ is defined as
\[
I_P:=\left(\max_{x\in P} f_1(x), \cdots, \max_{x\in P} f_m(x)\right).
\]
\end{definition}

This paper focuses on a bi-objective optimization problem in recommender systems, where accuracy and diversity serve as the two objective functions.

The first objective function, $f_1$, measures the accuracy of the recommendation list. It is defined as \citep{Chai21,Lacerda17,Wang16}:
\[
f_1 (R)= \frac{\sum_{i \in R, j \in P_u} S(i, j)}{|R|},
\]
where $R$ is the recommended list for user $u$, and $S(i, j)$ represents the similarity between items $i$ and $j$. The set $P_u$ refers to items rated by user $u$, and $|R|$ is the number of items in the recommended list. Cosine similarity is used to compute $S(i, j)$ \citep{Sarwar01}:
\begin{equation}\label{eq:predict:cosine}
S(i, j) := \frac{\langle R_i, R_j \rangle}{\| R_i \| \cdot \| R_j \|},
\end{equation}
where $R_i$ and $R_j$ represent the rating vectors corresponding to items $i$ and $j$, respectively. Additionally, $\langle \cdot, \cdot\rangle$ denotes the inner product, and $\Vert\cdot\Vert$ indicates the norm of a vector.

The second objective function, $f_2$, measures the diversity of the recommendation list. It is defined as \citep{Chai21,Wang16}:
\[
f_2 (R) = \frac{\sum_{i \in R} \sum_{j \in R, i \neq j} (1 - S(i, j))}{|R| \times (|R| - 1)}.
\]
Here, a lower similarity between items implies higher diversity in the recommended list.

For any user $u$, when items in the recommended list $R$ are highly similar, the accuracy tends to increase but at the cost of reduced diversity. This trade-off establishes accuracy and diversity as conflicting objectives in recommender systems.

The goal of this work is to achieve both high accuracy and diversity, effectively maximizing $f_1$ and $f_2$. The recommender system is formulated as a multi-objective optimization problem:
\[
\max_{R\in L_s} f(R) = \left(f_1(R), f_2(R)\right),
\]
where $L_s$ denotes the set of all sublists of the top-$k$ candidate list with size $s$.

\subsection{Prediction of unrated items}
In the first stage of HiMARS, we extract a top-$k$ candidate list for the target user. The data is represented as URM, where rows correspond to users and columns to items, and the $ij$-th component of matrix, i.e., $r_{i,j}$, denotes the rating that the $i$-th user gives to the $j$-th item. In practice, many items remain unrated by users. To generate a top-$k$ recommendation candidate list for a given user, it is necessary to predict the ratings for these unrated items. In this paper, we use the weighted sum approach \citep{Sarwar01} to predict the rating $R_{u,i}$ for user $u$ and item $i$, defined as follows:
\begin{equation}\label{eq:predict:weighted}
R_{u,i} := \frac{\sum_{j \in N}S(i,j) \times R_{u,j}}{\sum_{j \in N} |S(i,j)|},
\end{equation}
where $N\subseteq P_u$ represents the set of items $j$ most similar to item $i$ that have been rated by user $u$, and $R_{u,j}$ is the rating given by user $u$ to item $j$.

\subsection{HANv1 and HANv2}
These two algorithms are inspired by AMOSA \citep{Bandyopadhyay} and NSGA-II \citep{Deb}. Below, we highlight key aspects of AMOSA and NSGA-II, followed by an introduction to HANv1 and HANv2.

\subsubsection{AMOSA}
In AMOSA, we initialize an archive set, \texttt{Arc}, which can be done using a hill-climbing technique \citep{Bandyopadhyay} or by solving a weighted sum method to find some elements of the Pareto frontier. After initialization, a reference list, \texttt{current\_pt}, is selected and remains fixed throughout the algorithm's execution. New lists (\texttt{new\_pt}) are generated by perturbing \texttt{current\_pt} and are then compared with the elements in \texttt{Arc} based on their dominance relations.
Unlike many multi-objective algorithms, AMOSA allows accepting a dominated list with a probability that decreases as the algorithm progresses and depends on the amount of domination and a temperature parameter \texttt{$\tau$}. Specifically, the fewer lists that dominate a candidate list, the higher its acceptance probability. The archive size is controlled by two thresholds: the Hard Limit (HL), the maximum size allowed at termination, and the Soft Limit (SL), which triggers clustering when the archive size exceeds it. Clustering reduces the archive size to HL.

The acceptance probability of dominated lists is governed by parameters \texttt{$\tau$} and $\texttt{$\alpha$}<1$. The domination amount between two lists $a$ and $b$ is calculated as:
\[
\Delta dom_{a,b} = \prod_{\stackrel{i=1}{f_i(a)\neq f_i(b)}}^{2} \frac{\vert f_i(a) - f_i(b) \vert}{r_i},
\]
where $r_i$ is the range of the $i$th objective function. The process of creating new lists is described in Algorithm \ref{alg:new-list} where 
the reference set \texttt{PF} for AMOSA is \texttt{Arc}.

\subsubsection{NSGA-II}
\label{subsec:sngaii}
In NSGA-II, we initialize a population set, \texttt{Pop}, of top-$s$ recommendation lists by randomly sampling from the top-$k$ list generated in the first stage. The population size is determined by the parameter \texttt{pop\_size}. We then apply crossover on \texttt{Pop} to create $2 \times \left[\frac{\texttt{pop\_size} \times \texttt{pc}}{2}\right]$ offspring, where $[\cdot]$ denotes rounding to the nearest integer and \texttt{pc} represents the crossover probability. Mutation is subsequently applied, generating $\left[\texttt{pop\_size} \times \texttt{pm}\right]$ offspring, where \texttt{pm} is the mutation probability.

For crossover, we perform $\left[\frac{\texttt{pop\_size} \times \texttt{pc}}{2}\right]$ iterations. In each iteration, we randomly choose two elements of \texttt{Pop} as lists $L_1$ and $L_2$. Then, we choose a random number $\texttt{n\_cross} \in \{1, \cdots, s\}$. We divide $L_1$ and $L_2$ into two parts, denoting them as $L_{i,1}$ and $L_{i,2}$ for $i=1, 2$. We put the first \texttt{n\_cross} elements of $L_1$ into $L_{1,1}$ and the rest into $L_{1,2}$, and do the same for the $L_2$ list. 
We then create two new lists, $L_1^{\text{new}}$ and $L_2^{\text{new}}$, as follows: $L_1^{\text{new}} = [L_{1,1}, L_{2,2}]$ and $L_2^{\text{new}} = [L_{2,1}, L_{1,2}]$. In this process, some items may appear in both $L_{1,1}$ and $L_{2,2}$. To address this, we replace any duplicate items in $L_{2,2}$ with random items from $L_{2,1}$ that do not appear in $L_{1,1}$, ensuring all items in $L_1^{\text{new}}$ are unique. We apply a similar procedure for $L_2^{\text{new}}$. All new lists are stored in the set \texttt{Cp}, which holds the crossover offspring. The process of creating new lists via this approach is described in \citep{Chai21}.

For mutation, we perform $\left[\texttt{pop\_size} \times \texttt{pm}\right]$ iterations. In each iteration, we generate a random number and compare it with \texttt{pm}. If the random number is greater than \texttt{pm}, we randomly choose an element from \texttt{Cp} and, without any changes, add it to the set \texttt{Mp}, which stores the mutation offspring. Otherwise, we randomly select a list $L$ from \texttt{Cp}, randomly choose an item from $L$, and replace it with a random item from the top-$k$ list that has not previously appeared in $L$. The new list is then added to \texttt{Mp}.

\subsubsection{HANv1}
Although AMOSA is efficient, its initialization phase presents a challenge, as constructing the initial \texttt{Arc} requires additional effort. In the first version of HAN, we simplify this by generating \texttt{Arc} in each iteration without extra preprocessing, relying solely on the current \texttt{Pop} obtained through the NSGA-II process. Furthermore, \texttt{current\_pt} is not fixed throughout the algorithm's execution. We initialize \texttt{Pop} with top-$s$ recommendation lists, randomly sampled from the top-$k$ list. Crossover and mutation are applied as in NSGA-II. After non-dominated sorting and crowding distance calculations, we extract the first Pareto frontier into \texttt{Arc\_temp}. A list is then randomly selected from \texttt{Arc\_temp} as \texttt{current\_pt}, and we set \texttt{Arc} = \texttt{Nlists(current\_pt, Pop, $\tau$)} using Algorithm \ref{alg:new-list}. Finally, \texttt{Pop} and \texttt{Arc} are combined, as outlined in Algorithm \ref{alg:HANv1}.

\subsubsection{HANv2}
The second version of HAN, introduced by \cite{Lotfian23} and adapted for multi-objective recommender systems in our study, requires the initialization of \texttt{Arc} and fixing \texttt{current\_pt}. In this version, \texttt{Arc} is updated independently of \texttt{Pop}. After performing crossover and mutation, as in NSGA-II, the offspring are merged with \texttt{Pop} and \texttt{Arc}. The complete procedure is presented in Algorithm \ref{alg:HANv2}.

\begin{algorithm}[!h]
\caption{HANv1 algorithm}\label{alg:HANv1}
\begin{algorithmic}[1]
\Require  \texttt{Max\_Iter}, \texttt{SL}, \texttt{HL}, \texttt{$\alpha$}, \texttt{$\tau$}
\State \textbf{Initialization} $\texttt{Pop}=\{\text{initial population of size \texttt{HL}}\}$, $i=0$.
\While{$i\leq\texttt{Max\_Iter}$}
\State Do crossover on \texttt{Pop} and store it in \texttt{Cp}.
\State Do mutation on \texttt{Cp} and store it in \texttt{Mp}.
\State $\texttt{Pop}= \texttt{Cp}\cup\texttt{Mp}\cup \texttt{Pop}$.
\State  Sort \texttt{Pop} using non-dominated sorting and crowding distance. Put the first Pareto frontier in \texttt{Arc\_temp}.
\State  Select $\text{list}\in\texttt{Arc\_temp}$ randomly as \texttt{current\_pt} and set
\texttt{Arc}$=$\texttt{Nlists(current\_pt, Pop, $\tau$)} (Algorithm \ref{alg:new-list}). 
\State  Set $\texttt{$\tau$}=\texttt{$\alpha$}\times \texttt{$\tau$}$ and $\texttt{Pop}=\texttt{Pop}\cup\texttt{Arc}$.
\State  Sort \texttt{Pop} using non-dominated sorting and crowding distance.
Keep the first \texttt{HL} numbers of lists corresponding to the first Pareto frontier in \texttt{Pop}  and remove rest.
\State i = i+1
\EndWhile
\end{algorithmic}
\end{algorithm}

\begin{algorithm}[!h]
\caption{HANv2 algorithm}\label{alg:HANv2}
\begin{algorithmic}[1]
\Require  \texttt{Max\_Iter}, \texttt{SL}, \texttt{HL}, \texttt{$\alpha$}, \texttt{$\tau$}
\State \textbf{Initialization} $\texttt{Pop}=\{\text{initial population of size \texttt{HL}}\}$, \texttt{Arc}, $i=0$.
Select $\text{list}\in\texttt{Arc}$ randomly as \texttt{current\_pt}.
\While{$i\leq\texttt{Max\_Iter}$}
\State Set \texttt{Arc}$=$\texttt{Nlists(current\_pt, Arc, $\tau$)} (Algorithm \ref{alg:new-list}) and  $\texttt{$\tau$}=\texttt{$\alpha$}\times \texttt{$\tau$}$.
\State Do crossover on \texttt{Pop} and store it in \texttt{Cp}.
\State Do mutation on \texttt{Cp} and store it in \texttt{Mp}.
\State $\texttt{Pop}=\texttt{Cp} \cup \texttt{Mp}\cup \texttt{Pop}\cup \texttt{Arc}$.
\State  Sort \texttt{Pop} using non-dominated sorting and crowding distance.
Keep the first \texttt{HL} numbers of lists corresponding to the first Pareto frontier in \texttt{Pop}  and remove rest.
\State i = i+1
\EndWhile
\end{algorithmic}
\end{algorithm}

\begin{algorithm}[!h]
\caption{\texttt{Nlists(current\_pt, PF, $\tau$)}.}\label{alg:new-list}
\scalebox{0.86}{ 
\parbox{\linewidth}{
\begin{algorithmic}[1]
\Require \texttt{current\_pt}, \texttt{PF} (reference set), \texttt{$\tau$}, \texttt{SL}, \texttt{HL}.
\State $\texttt{NI},I_m=\texttt{NIitems(current\_pt)}$ (Algorithm \ref{alg:non_included_items}).
\For{$i\in I_m$}
\State  \texttt{new\_pt}$=$ replace $i$-th item of \texttt{current\_pt} by a random item in \texttt{NI}($i_{\texttt{NI}}$).
\If {$f(\texttt{current\_pt}) \gneqq  f(\texttt{new\_pt})$}
\State set $t=\left\vert\{\text{list}\in\texttt{PF}\mid \text{list}\gvertneqq\texttt{new\_pt}\}\right\vert$,
\State set $\Delta dom_{avg}=\frac{\left(\sum_{j=1}^{t} \Delta dom_{j, \texttt{new\_pt}}\right)+\Delta dom_{\texttt{current\_pt}, \texttt{new\_pt}}}{t+1},$
  \If{$\texttt{rand}<\frac{1}{1+\exp(\Delta dom_{avg}\times \texttt{$\tau$})}$}
 \State set $\texttt{current\_pt}= \texttt{new\_pt}$.
       \EndIf
  \ElsIf{\texttt{current\_pt} $\nleq\ngeq$ \texttt{new\_pt} }
          \If{\texttt{new\_pt} is dominated by $t$  lists in \texttt{PF}}
          \If{$\texttt{rand}<$ {\tiny$\frac{1}{1+\exp\left(\frac{\left(\sum_{j=1}^{t} \Delta dom_{j, \texttt{new\_pt}}\right)}{t}\times \texttt{$\tau$}\right)}$}}
           \State set $\texttt{current\_pt}= \texttt{new\_pt}$.
       \EndIf
         \ElsIf {\texttt{new\_pt} is non-dominated w.r.t \texttt{PF}} 
         \State set $\texttt{current\_pt}=\texttt{new\_pt}$ and add it to \texttt{PF}. 
         \If {$|\texttt{PF}|>|\texttt{SL}|$}\State use clustering procedure  to thin \texttt{PF} to \texttt{HL}.\EndIf
         \ElsIf {\texttt{new\_pt} dominated $t$ lists in \texttt{PF}} 
         \State set $\texttt{current\_pt}= \texttt{new\_pt}$ and add it to \texttt{PF},
         then remove $t$ lists from \texttt{PF}.
       \EndIf
    \ElsIf{$f(\texttt{new\_pt})\gneqq f(\texttt{current\_pt})$}
       \If{\texttt{new\_pt} is dominated by $t$ lists in \texttt{PF}}
          \State set $\Delta dom_{min}=\min\{\Delta dom_{\texttt{new-pt},t}\mid t\in \text{list}\}$.
            \If{$\texttt{rand}<\frac{1}{1+\exp(-\Delta dom_{min})}$}
             \State set \texttt{current\_pt}$=$ the list of \texttt{PF} corresponded to $\Delta dom_{min}$.
             \EndIf 
       \Else \State $\texttt{current\_pt}= \texttt{new\_pt}$.
       \EndIf
       \If {\texttt{new\_pt} is non-dominated w.r.t the lists in \texttt{PF}}
        \State set $\texttt{current\_pt}= \texttt{new\_pt}$ and add it to \texttt{PF}. 
        \If {\texttt{current\_pt} is in \texttt{PF}}\State remove it. \ElsIf {$|\texttt{PF}|> |\texttt{SL}|$}\State use clustering procedure  to thin \texttt{PF} to \texttt{HL}.\EndIf
       \ElsIf {\texttt{new\_pt} dominates $t$ other lists in \texttt{PF}}
        \State $\texttt{current\_pt}= \texttt{new\_pt}$  and add it to \texttt{PF},
         then remove $t$ lists from \texttt{PF}.
       \EndIf
\EndIf
\State Remove $i_{\texttt{NI}}$ from \texttt{NI}.
\EndFor
\State Return \texttt{PF}.
\end{algorithmic}
}}
\end{algorithm}

\begin{algorithm}[!h]
\caption{\texttt{NIitems}(\texttt{current\_pt}).}\label{alg:non_included_items}
\begin{algorithmic}[1]
\Require \texttt{current\_pt}.
\State Set $\texttt{NI}=\{\text{items in Top-$k$ lists}\}\setminus \{\text{items in}~\texttt{current\_pt}\}$ and compute $m=\min\{s, |\texttt{NI}|\}$.
\If {$m=s$}
  \State set $I_m=\{1, \dots, s\}$
  \Else
  \State set $I_m=\left\{j\in \{1, \dots, s\}\mid j~\text{randomly is selected}\right\}$ and the size of $I_m= |\texttt{NI}|$.
\EndIf
\State Return \texttt{NI} and $I_m$.
\end{algorithmic}
\end{algorithm}

\subsection{HANIv1 and HANIv2}
These two algorithms are inspired by AMOSA \citep{Bandyopadhyay} and NNIA \citep{Gong08}. Below, we highlight key aspects of NNIA, followed by an introduction to HANIv1 and HANIv2.

\subsubsection{NNIA}
In NNIA \citep{Gong08}, inspired by the immune system, we initialize a population set, \texttt{$B_0$}, of top-$s$ recommendation lists by randomly sampling from the top-$k$ list generated in the first stage. The population size is determined by the parameter \texttt{nd}. Non-dominated lists from \texttt{$B_0$} are stored in \texttt{D}, which is then sorted by crowding distance. We retain the top \texttt{nd} lists in \texttt{D} and remove the rest. The first \texttt{na} lists in \texttt{D} form the active population set \texttt{A}, and a clone set \texttt{C} of size \texttt{nc} is generated from \texttt{A}, as explained in \citep{Gong08}.

For crossover, we create $2\times \texttt{nc}$ elements by performing $\texttt{nc}$ iterations. In the $i$th iteration, we choose a random list from \texttt{A} and the $i$th list from \texttt{C}. Then, we create two new lists as described in Subsection \ref{subsec:sngaii}. All new lists are placed in the set \texttt{Cp}, which holds the crossover offspring.

For mutation, we create $2\times \texttt{nc}$ elements by performing $2\times \texttt{nc}$ iterations. The process of creating a new list is similar to the mutation process described in Subsection \ref{subsec:sngaii}.

\subsubsection{HANIv1 and HANIv2}
The processes in HANIv1 and HANIv2 are analogous to HANv1 and HANv2, respectively, with NNIA replacing NSGA-II. See Algorithm \ref{alg:HANIv1} for HANIv1 and Algorithm \ref{alg:HANIv2} for HANIv2.

\begin{algorithm}[!h]
\caption{HANIv1 algorithm}\label{alg:HANIv1}
\begin{algorithmic}[1]
\Require  \texttt{Max\_Iter}, \texttt{nd}, \texttt{na}, \texttt{nc}
\State \textbf{Initialization} $\texttt{$B_0$}=\{\text{initial population of size \texttt{nd}}\}$, $i=0$.
\While{$i\leq\texttt{Max\_Iter}$}
\State $\texttt{D}= \text{non-dominated lists of}~\texttt{$B_0$}$.
\State Sort \texttt{D} in descending order of crowding distance. Keep the first \texttt{nd} numbers of lists in \texttt{D} and remove rest.
\State $\texttt{A}=$ first \texttt{na} lists of \texttt{D}
\State $\texttt{C}= $ clone of \texttt{A} with size  \texttt{nc}.
\State Do crossover between \texttt{C} and \texttt{A} and store it in \texttt{CT}.
\State Do mutation on \texttt{CT} and store it in \texttt{Ct}.
\State  Select $\text{list}\in\texttt{A}$ randomly as \texttt{current\_pt} and set
\texttt{Arc}$=$\texttt{Nlists(current\_pt, \texttt{$B_i$}, $\tau$)} (Algorithm \ref{alg:new-list}). 
\State  Set $\texttt{$B_i$}=\texttt{Ct}\cup\texttt{D}\cup\texttt{Arc}$.
\State i = i+1
\EndWhile
\end{algorithmic}
\end{algorithm}

\begin{algorithm}[!h]
\caption{HANIv2 algorithm}\label{alg:HANIv2}
\begin{algorithmic}[1]
\Require  \texttt{Max\_Iter}, \texttt{SL}, \texttt{HL}, \texttt{$\alpha$}, \texttt{$\tau$}, \texttt{nd}, \texttt{na}, \texttt{nc}
\State \textbf{Initialization} $\texttt{$B_0$}=\{\text{initial population of size \texttt{nd}}\}$, \texttt{Arc}, $i=0$.
Select $\text{list}\in\texttt{Arc}$ randomly as \texttt{current\_pt}.
\While{$i\leq\texttt{Max\_Iter}$}
\State Set \texttt{Arc}$=$\texttt{Nlists(current\_pt, Arc, $\tau$)} (Algorithm \ref{alg:new-list}) and  $\texttt{$\tau$}=\texttt{$\alpha$}\times \texttt{$\tau$}$.
\State $\texttt{D}= \text{non-dominated lists of}~\texttt{$B_0$}$.
\State Sort \texttt{D} in descending order of crowding distance. Keep the first \texttt{nd} numbers of lists in \texttt{D} and remove rest.
\State $\texttt{A}=$ first \texttt{na} lists of \texttt{D}
\State $\texttt{C}= $ clone of \texttt{A} with size  \texttt{nc}.
\State Do crossover between \texttt{C} and \texttt{A} and store it in \texttt{CT}.
\State Do mutation on \texttt{CT} and store it in \texttt{Ct}.
\State  Set $\texttt{$B_i$}=\texttt{Ct}\cup\texttt{D}\cup\texttt{Arc}$.
\State i = i+1
\EndWhile
\end{algorithmic}
\end{algorithm}

\subsection{Selection of a personalized optimal top-$s$ list}
The selection of a final optimal top-$s$ list for recommendation is based on its proximity to an ideal solution. In this approach, we follow the process below. If 
$\mathcal{P}=\{P_1, \cdots, P_M\}$ is the resulting Pareto frontier generated by an algorithm, we first 
scale the values of the objective functions related to this frontier. Then, we 
create the ideal point by assigning the maximum value of the first scaled objective function to its first component and the maximum value of the second scaled objective function to its second component. Next, we calculate the Euclidean distance of each Pareto solution $P_i$ to the ideal point:
\[
d_i = \sqrt{\sum_{j=1}^{2} (f_j(P_i) - \text{ideal}_j)^2},
\]
where $f_j(P_i)$ is the value of the $j$th objective function for solution $P_i$ and $\text{ideal}_j$ is the $j$th component of the ideal point. The solution $P^*$ that minimizes this distance, i.e., 
\[
P^* = \arg\min_{P_i \in \mathcal{P}} d_i,
\]
is selected as the final personalized top-$s$ list for recommendation to target user.

\section{Evaluation of proposed methods}
\label{sec:evaluation}
To evaluate the proposed algorithms in comparison with existing methods in the literature, this paper utilizes two categories of metrics. The first category
assesses the quality of the recommendation lists while the second category evaluates the quality of the Pareto frontiers generated by the multi-objective algorithms.

\subsection{Evaluating  the recommendation quality}
Among the various metrics discussed in the literature for evaluating recommender systems, this paper focuses on three key metrics: accuracy, diversity, and novelty \citep{Avazpour14,Geng15,Wang16}.

\begin{itemize}
\item \textbf{Accuracy:}  
Accuracy measures the proportion of relevant items included in the recommendation list for a given user. It is defined as:
\[
P(R)=\frac{\left|R\cap T\right|}{\left|R\right|},
\]
where $R$ denotes the recommendation list and $T$ represents the items rated by the target user in the testing data set, filtered by a lower bound threshold. In this paper, the threshold considered to be 3. Here, $|\cdot|$ indicates the size of a list.
A higher $P(R)$ indicates more accurate recommendations.

\item \textbf{Diversity:}  
To assess the variety of recommendations, this paper employs intra-user diversity, computed using the following formula:
\[
D(R)=\frac{1}{\left|R\right|\times \left(\left|R\right|-1\right)}\sum_{i\ne j}S(i,j),
\]
where $S(i, j)$ measures the similarity between items $i$ and $j$ in the recommendation list $R$, and $\left|R\right|$ is the size of the recommendation list. A lower $D(R)$ value indicates higher diversity.

\item \textbf{Novelty:}  
Novelty reflects how uncommon or unexpected the recommended items are. It is defined as:
\[
N(R)=\frac{1}{\left|R\right|}\sum_{i=1}^{\left|R\right|}{J}_{i},
\]
where $J_i$ represents the number of users who have rated item $i$. Lower values of $N(R)$ indicate higher novelty.
\end{itemize}

\subsection{Evaluating  the quality of  Pareto frontiers}
\label{subsec:evParfront}

Note that a recommender system might show favorable results for accuracy and diversity individually but fail to provide balanced recommendation lists across both objectives. Metrics that assess the quality of Pareto frontiers are crucial for evaluating aspects such as the uniformity of the Pareto frontier's distribution and the distance between the ideal point and the frontier.
Considering problem \ref{eq3}, we evaluate the performance of multi-objective optimization algorithms using four metrics that assess the diversity and coverage of the Pareto frontier, as outlined by \citep{Maghsoudlou16,Zitzler98}.

\begin{itemize}
    \item \textbf{Spacing Metric (SM):}   
    To assess the uniformity of the distribution along the Pareto frontier, we employ the following formula:
\[
{\rm SM} = \frac{\sum_{i=1}^{n-1} d_i }{(n - 1)},
\]
where $d_i$ denotes the distance between consecutive non-dominated solutions on the Pareto frontier and $n$ signifies the total number of solutions in the Pareto frontier.

    \item \textbf{Mean Ideal Distance (MID):} 
    To quantify the distance between an ideal point and the Pareto frontier, we utilize the following formula:
\[
{\rm MID} = \frac{\sum_{i=1}^{n} \left[ \left( \frac{f_{1i} -f_{1total}^{\max}}{f_{1total}^{\max} - f_{1total}^{\min}} \right)^2 + \left( \frac{f_{2i} - f_{2total}^{\max}}{f_{2total}^{\max} - f_{2total}^{\min}} \right)^2 \right]^{1/2}}{n},
\]
where $f_{1i}$ and $f_{2i}$ represent the first and second objective values for the $i$th Pareto point, respectively. The minimum and maximum values of the $j$th objective function across all Pareto solutions are denoted by $f_{jtotal}^{\min}$ and $f_{jtotal}^{\max}$, for $j = 1, 2$.

    \item \textbf{Diversification Metric (DM):} This metric evaluates the extension of the Pareto frontier, calculated as:
    \[
     {\rm DM} = \left[ \sum_{i=1}^{m} \left( \min f_i - \max f_i \right)^2 \right]^{1/2},
    \]
    where $f_i$ is the value of the $i$th objective function.

    \item \textbf{Spread of Non-dominance Solution (SNS):} This metric measures the spread of the Pareto frontier:
    \[
     {\rm SNS} = \left[ \frac{\sum_{i=1}^{n} ({\rm MID} - C_i)^2}{n - 1} \right]^{1/2},
    \]
    where $C_i = \left[ f_{1i}^2 + f_{2i}^2 \right]^{1/2}$.
\end{itemize}
An algorithm with lower values of SM and MID, alongside higher values of DM and SNS, is typically better.

It is important to note that if an algorithm generates only two Pareto solutions, one maximizing accuracy and the other maximizing diversity, then the DM value may be large. However, this outcome may reflect the algorithm's inability to generate balanced solutions, and the limited number of solutions also restricts the diversity of proposed lists. A similar issue arises if the algorithm creates numerous Pareto solutions that are clustered into two groups: one group concentrated near high accuracy and the other near high diversity. 
Additionally, if all points are clustered around the middle of the Pareto frontier, the MID metric will display a low value. However, such an outcome may indicate that the algorithm may fail to offer lists with sufficiently high accuracy or diversity, limiting its practical utility. 
Thus, when using DM, MID, and similar metrics to compare multi-objective algorithms, it is crucial to employ a multi-criteria decision analysis (MCDA) method that assigns appropriate weights to these metrics and ranks algorithms accordingly. This approach ensures a more holistic evaluation of algorithm performance across multiple objectives.

In our approach, we use the Technique for Order of Preference by Similarity to Ideal Solution (TOPSIS) \citep{Hwang1981} to rank the algorithms based on four performance metrics. Given the importance of uniformity and diversity of the Pareto frontier, as well as the high sensitivity of the DM and MID metrics to the number and distribution of Pareto solutions, we assign weights of $w_{\text{SM}}=w_{\text{SNS}}=0.33$ and $w_{\text{MID}}=w_{\text{DM}}=0.17$ to these metrics based on the AHP method \citep{Saaty08}.

To implement TOPSIS, we first construct a decision matrix containing the values of these metrics for the target user. We then create a weighted normalized decision matrix, with entries defined as:
$\nu _{ij} =r_{ij}\times \omega$, where
\[
 r_{ij} =\frac{x_{ij}}{\left( \sum _{i=1}^4 x_{ij}^2\right) ^{1/2}}, 
\]
and $r_{ij}$ denotes the components of the normalized decision matrix, with $x_{ij}$ representing the components of the original decision matrix. The weight matrix $\omega$ is diagonal, with $w_{\text{SM}}$, $w_{\text{SNS}}$, $w_{\text{MID}}$, and $w_{\text{DM}}$ along the diagonal, and other entries set to zero.

The positive and negative ideal solutions are then defined as:
\begin{align*} 
A^+&=[\nu _1^+,\cdots, \nu _4^+] =\left\{ \left( \max _{j}\nu _{ij}\vert i \in I''\right) ,\left( \min _{j}\nu _{ij}\vert i \in I' \right) \right\} ,\\
A^-&=[\nu _1^- ,\cdots, \nu _4^-] =\left\{ \left( \min _{j}\nu _{ij}\vert i \in I''\right) ,\left( \max _{j}\nu _{ij}\vert i \in I' \right) \right\} , 
\end{align*}
where $I'$ represents criteria with a negative impact (in this case, SM and MID), and $I''$ represents criteria with a positive impact (DM and SNS).

We then calculate the separation distance of the $i$th algorithm from the positive and negative ideal solutions as:
\begin{equation*} 
d_i^+=\left[ \sum _{j=1}^{4}(\nu _{ij}-\nu _j^+)^2\right] ^{1/2},\quad d_i^-=\left[ \sum _{j=1}^{4}(\nu _{ij}-\nu _j^-)^2\right] ^{1/2}. 
\end{equation*}
Finally, the relative closeness of the $i$th algorithm to the ideal solution is computed as:
\[
\textrm{CLO}_i=\frac{d_i^-}{d_i^-+d_i^+}. 
\]
The algorithm with the highest $\text{CLO}_i$ value is considered the best.

\subsection{Experimental results}

\subsubsection{Dataset}
To validate the proposed methods, experiments were conducted on two datasets: the MovieLens dataset, containing 6,040 users and 3,706 movies\footnote{http://grouplens.org/datasets/movielens}, and the ModCloth dataset, consisting of 44,784 users and 1,020 items\footnote{https://cseweb.ucsd.edu/~jmcauley/datasets.html}; see also \cite{Misra18}. The sparsity of the MovieLens dataset is 82.13\%, while the ModCloth dataset exhibits a considerably higher sparsity of 99.34\%. We compared the performance of the proposed hybrid algorithms, i.e., HANv1, HANv2, HANIv1, and HANIv2, with ICF, NNIA, AMOSA, and NSGA-II for a subset of ten users. For MovieLens, we selected users 3411-3420, following prior studies \citep{Chai21, Zou21}, while for ModCloth, users 620-629 were randomly chosen.

For both datasets, we randomly divided ratings into an 80\% training set and a 20\% test set. The training set was used to build the recommender systems, while the test set served for evaluation. The high sparsity of these datasets, particularly the ModCloth dataset, presents a significant challenge, as the sparsity of user-item interactions makes generating diverse and novel recommendations more difficult.

\subsubsection{Methods and parameter adjustment}
We compare the proposed hybrid methods with ICF, NNIA, AMOSA, and NSGA-II. All algorithms were implemented in Python 3.11.3 and executed on a laptop equipped with a 12th Gen Intel$\circledR$~Core$^{\text{TM}}$ i7-12800H CPU (1.80 GHz) and 16 GB of RAM. The results are based on 20 simulations, and the parameters for the experiments are defined as follows:
\begin{itemize}
\item \textbf{General Parameters:}
\begin{itemize}
  \item $\texttt{Max\_Iter}=200$: Number of iterations for multi-objective algorithms.
  \item Top-$k=100$: Size of the initial top-$k$ recommendation list.
  \item Top-$s=10$: Size of the final top-$s$ recommendation list.
  \item $N= 20$: Number of most similar items for rating prediction using \eqref{eq:predict:weighted}.
\end{itemize} 

\item \textbf{ICF:} For Item-Based Collaborative Filtering (ICF) \citep{Sarwar01}, Adjusted Cosine Similarity \cite[Subsection 3.1.3]{Sarwar01} is used to calculate item similarity, and the Weighted Sum method, as defined in \eqref{eq:predict:weighted}, is applied for rating prediction. ICF generates the initial top-$k$ list in the first stage of HiMARS, from which the optimal top-$s$ list is selected.
    
\item \textbf{NNIA:} Parameters are set to $\texttt{nd}=100$, $\texttt{na}=10$, $\texttt{nc}=40$, with a mutation probability $\texttt{pm}=0.1$.

\item \textbf{AMOSA:} Parameters include $\texttt{SL}=140$, $\texttt{HL}=100$, $\tau=1$, and $\alpha=0.9$.

\item \textbf{NSGA-II:} The population size is set to $\texttt{pop\_size}=100$, with a crossover probability $\texttt{pc}=0.7$ and mutation probability $\texttt{pm}=0.2$.

\item \textbf{HANv1} and \textbf{HANv2:} Parameters are aligned with those of AMOSA and NSGA-II.

\item \textbf{HANIv1} and \textbf{HANIv2:} Parameters follow those of AMOSA and NNIA.
\end{itemize}
For AMOSA, HANv2, and HANIv1, an initial archive \texttt{Arc} is generated using NNIA with $\texttt{Max\_Iter}=50$. The Pareto set obtained from this process is used as the initial archive in these algorithms. All of the parameters are obtained by trial and error.

\subsubsection{Accuracy}
\begin{figure}[!h]
\centering
\includegraphics[width=0.5\linewidth]{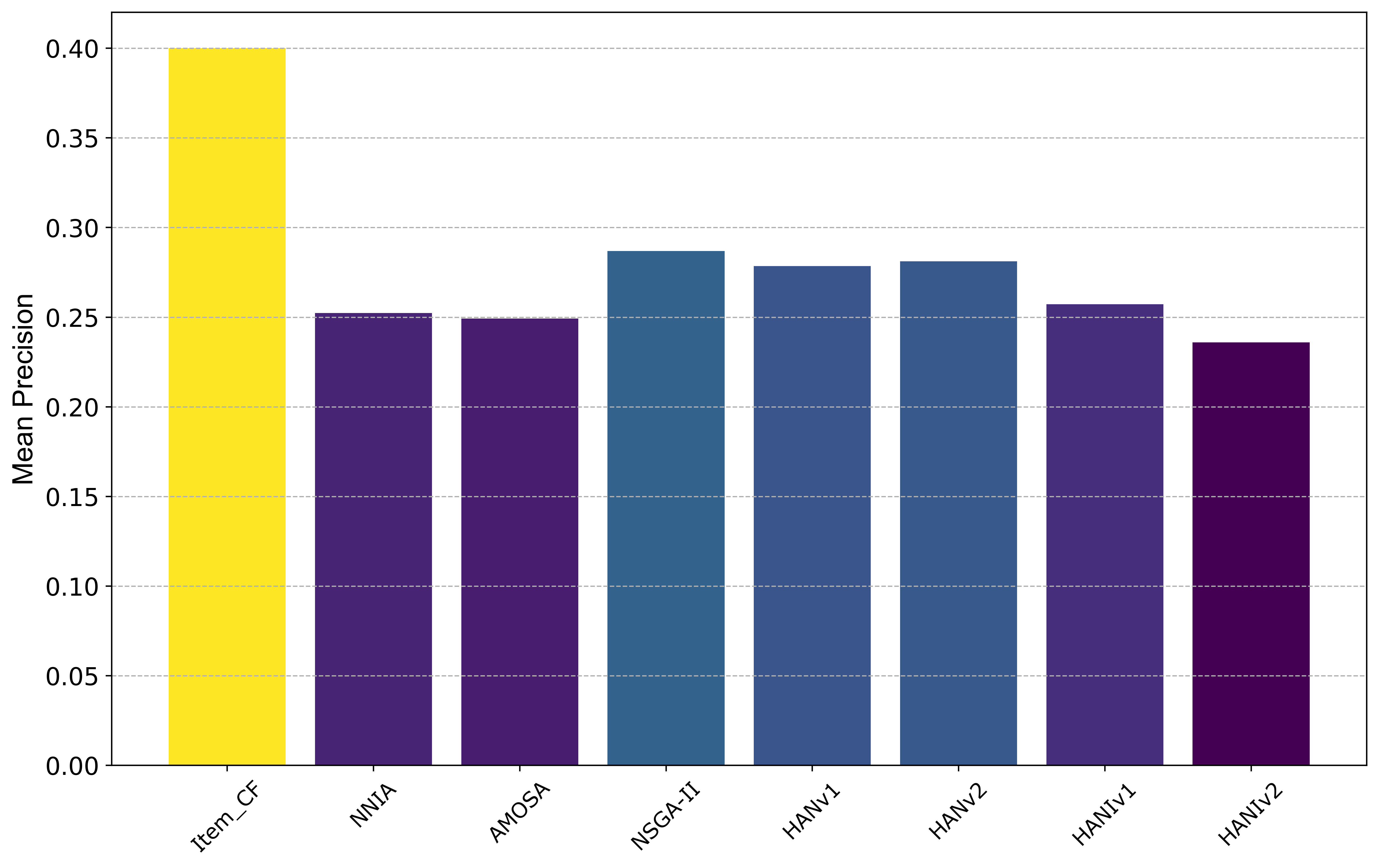}
\caption{Accuracy  comparisons on MovieLens.}\label{fig:precommovie}
\end{figure}

Table \ref{tab:Accuracy:Movie} presents the accuracy results of the recommendation lists on the MovieLens dataset. A higher value indicates a more accurate recommendation list. Since ICF primarily focuses on accuracy, it generally outperforms the multi-objective algorithms in mean accuracy for most users.
Among the multi-objective algorithms, HANv1 achieves the highest minimum values across most users, indicating its robustness in generating consistently accurate recommendations. For maximum accuracy values, AMOSA and HANv2 perform best overall, suggesting these algorithms excel at producing highly accurate recommendations.
NSGA-II, when considering mean accuracy, shows the best performance among the multi-objective algorithms. Figure \ref{fig:precommovie} visualizes the mean accuracy results across users, further highlighting ICF's strength in delivering accurate recommendations. Among the multi-objective methods, NSGA-II, HANv1, and HANv2 demonstrate high efficiency in achieving accurate recommendations.

\begin{figure}[!h]
\centering
\includegraphics[width=0.5 \linewidth]{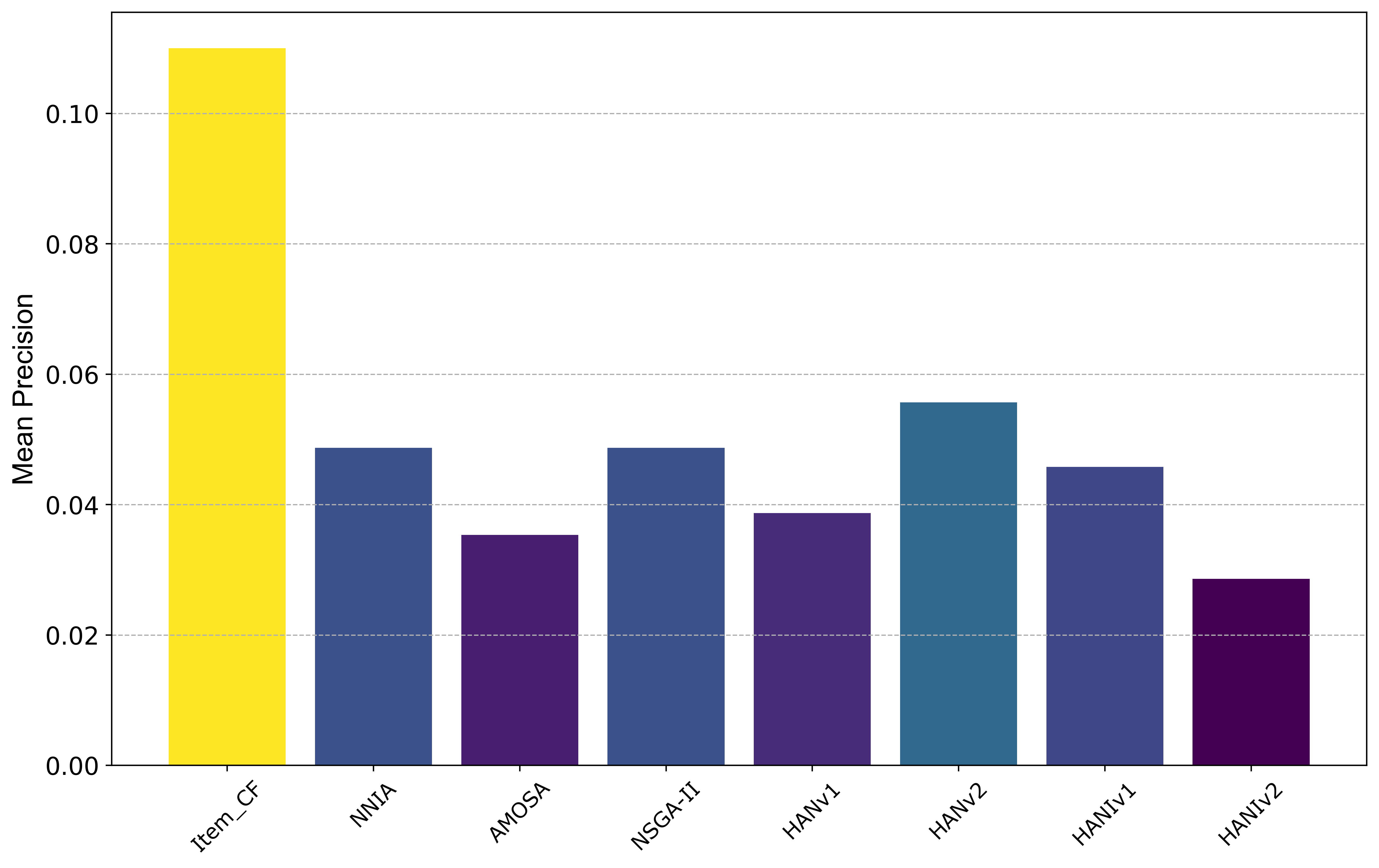}
\caption{Accuracy  comparisons on ModCloth.}\label{fig:precomModCloth}
\end{figure}

Table \ref{tab:Accuracy:ModCloth} presents the accuracy results on the ModCloth dataset. As with the MovieLens dataset, ICF performs best in terms of accuracy, benefiting from its primary focus on delivering accurate item recommendations. Given the high sparsity of the ModCloth dataset, high accuracy values across all algorithms were not anticipated.
Nevertheless, Figure \ref{fig:precomModCloth} illustrates that among the multi-objective optimization algorithms, HANv2 demonstrates promising accuracy, indicating it may be better suited for handling sparse data.

\begin{table*}[!h]
\caption{\label{tab:Accuracy:Movie} Accuracy on MovieLens.}
   \scalebox{0.4}{
    \begin{tabular*}{\linewidth}{@{\extracolsep{\fill}}cccccccccccccccccccccccccccccccc@{}}
\cmidrule(lr){1-30}
UserID & Item-CF &&& NNIA &  &&&NSGAII&&&  &AMOSA&&&  & HANv1&&&  & HANv2 &&&  & HANIv1 &&&  & HANIv2\\
\cmidrule(lr){4-6} \cmidrule(lr){8-10} \cmidrule(lr){12-14} \cmidrule(lr){16-18} \cmidrule(lr){20-22} \cmidrule(lr){24-26} \cmidrule(lr){28-30}
          &&&$\min$ & $\max$ & mean&& $\min$& $\max$& mean&& $\min$& $\max$& mean&& $\min$& $\max$& mean&& $\min$& $\max$& mean&& $\min$& $\max$& mean&& $\min$& $\max$& mean\\
\cmidrule(lr){1-30}
3411 & 0.5 && 0.08 & 0.4 & 0.27 && 0.1 & 0.6 & 0.37 && 0.04 & 0.6 & 0.32 && \textbf{0.22} & 0.5 & 0.36 &&
 0.1 & \textbf{\color{blue}{0.62}} & \textbf{\color{red}{0.38}} && 0.08 & 0.42 & 0.25 && 0.02 & 0.56 & 0.29 \\
 
3412 & 0.8 && 0.12 & 0.66 & \textbf{\color{red}{0.43}} && 0.12 & 0.68 & 0.4 && 0.06 & 0.64 & 0.33 && \textbf{0.18} & 0.58 & 0.36 && 0.06 & \textbf{\color{blue}{0.7}} & 0.37 && 0.16 & 0.62 & 0.39 && 0.08 & 0.66 & 0.41 \\

3413 & 0.4 && 0.02 & 0.22 & 0.11 && 0.02 & \textbf{\color{blue}{0.42}} & \textbf{\color{red}{0.22}} && 0.0 & 0.3 & 0.11 && \textbf{0.04} & 0.3 & 0.16 && 0.0 & 0.36 & 0.14 && 0.0 & 0.22 & 0.07 && 0.0 & 0.26 & 0.09 \\

3414 & 0.7 && 0.16 & 0.8 & 0.5 && 0.34 & 0.82 & \textbf{\color{red}{0.61}} && 0.1 & 0.84 & 0.51 && \textbf{0.36} & 0.8 & 0.57 && 0.18 & \textbf{\color{blue}{0.9}} & 0.6 && 0.14 & 0.78 & 0.46 && 0.12 & 0.82 & 0.38 \\

3415 & 0.3 && 0.1 & 0.34 & 0.22 && 0.0 & 0.42 & 0.18 && 0.0 & 0.38 & 0.16 && 0.08 & 0.36 & 0.23 
&& 0.06 & \textbf{\color{blue}{0.48}} & \textbf{\color{red}{0.26}} && \textbf{0.18 }& 0.3 & 0.24 && 0.0 & 0.4 & 0.17 \\

3416 & 0.6 && 0.04 & 0.6 & 0.24 && \textbf{0.14} & 0.58 & 0.32 && 0.0 & \textbf{\color{blue}{0.66}} & 0.32 
&& \textbf{0.14} & 0.5 & 0.31 && 0.04 & 0.6 & 0.29 && 0.02 & 0.58 & \textbf{\color{red}{0.36}} && 0.0 & \textbf{\color{blue}{0.66}} & 0.35 \\

3417 & 0.2 && 0.06 & 0.18 & 0.14 && 0.0 & 0.22 & 0.1 && 0.0 & \textbf{\color{blue}{0.36}} & 0.14 && 0.02 & 0.22 & 0.1 && 0.02 & 0.3 & 0.16 && \textbf{0.16} & 0.24 & \textbf{\color{red}{0.19}} && 0.04 & 0.34 & 0.14 \\

3418 & 0.1 && 0.08 & 0.3 & 0.19 && 0.08 & \textbf{\color{blue}{0.46}} & \textbf{\color{red}{0.26}} && 0.0 & 0.44 & 0.2 && 0.12 & 0.42 & 0.25 && 0.04 & 0.44 & 0.22 && \textbf{0.16} & 0.26 & 0.22 && 0.0 & 0.4 & 0.18 \\

3419 & 0.1 && \textbf{0.06} & 0.24 & 0.14 && 0.02 & 0.3 & \textbf{\color{red}{0.16}} 
&& 0.0 & \textbf{\color{blue}{0.38}} & \textbf{\color{red}{0.16}} && \textbf{0.06} & 0.3 & \textbf{\color{red}{0.16}} && 0.0 & 0.32 & 0.15 && 0.02 & 0.24 & 0.09 && 0.0 & 0.26 & 0.1 \\

3420 & 0.3 && 0.06 & 0.5 & 0.29 && 0.04 & 0.52 & 0.26 && 0.0 & \textbf{\color{blue}{0.58}} & 0.24 
&& \textbf{0.12} & 0.44 & 0.28 && 0.02 & 0.52 & 0.23 && 0.06 & 0.56 & \textbf{\color{red}{0.32}} && 0.0 & 0.56 & 0.26 \\
\cmidrule(lr){1-30}
\end{tabular*}}
\end{table*}

\begin{table*}[!h]
\caption{\label{tab:Accuracy:ModCloth} Accuracy on ModCloth.}
   \scalebox{0.405}{
    \begin{tabular*}{\linewidth}{@{\extracolsep{\fill}}cccccccccccccccccccccccccccccccc@{}}
\cmidrule(lr){1-30}
UserID & Item-CF &&& NNIA &  &&&NSGAII&&&  &AMOSA&&&  & HANv1&&&  & HANv2 &&&  & HANIv1 &&&  & HANIv2\\
\cmidrule(lr){4-6} \cmidrule(lr){8-10} \cmidrule(lr){12-14} \cmidrule(lr){16-18} \cmidrule(lr){20-22} \cmidrule(lr){24-26} \cmidrule(lr){28-30}
          &&&$\min$ & $\max$ & mean&& $\min$& $\max$& mean&& $\min$& $\max$& mean&& $\min$& $\max$& mean&& $\min$& $\max$& mean&& $\min$& $\max$& mean&& $\min$& $\max$& mean\\
\cmidrule(lr){1-30}
620 & 0.1 && 0.0 &  \textbf{\color{blue}{0.1}} & 0.0 && 0.0 & 0.0 & 0.0 && 0.0 &  \textbf{\color{blue}{0.1}} & \textbf{\color{red}{0.01}} && 0.0 & 0.0 & 0.0 
&& 0.0 & 0.0 & 0.0 && 0.0 & 0.0 & 0.0 && 0.0 &  \textbf{\color{blue}{0.1}} &  \textbf{\color{red}{0.01}} \\

621 & 0.0 && 0.0 & 0.0 & 0.0 && 0.0 & 0.0 & 0.0 && 0.0 & 0.0 & 0.0 && 0.0 & 0.0 & 0.0 && 0.0 & 0.0 & 0.0 && 0.0 & 0.0 & 0.0 && 0.0 & 0.0 & 0.0 \\

622 & 0.0 && 0.0 & 0.1 & 0.04 && 0.0 & 0.1 & 0.05 && 0.0 & 0.1 & 0.03 && 0.0 & 0.1 & 0.05 && 0.0 &  \textbf{\color{blue}{0.2}} &  \textbf{\color{red}{0.11}} && 0.0 & 0.0 & 0.0 && 0.0 & 0.1 & 0.0 \\

623 &0.3 && 0.0 & 0.1 & 0.04 && \textbf{0.1} & \textbf{\color{blue}{0.3}} &  \textbf{\color{red}{0.18}} && 0.0 & \textbf{\color{blue}{0.3}} & 0.07 
&&\textbf{0.1} & 0.2 & 0.14 && 0.0 & \textbf{\color{blue}{0.3}} & 0.12 && \textbf{0.1} & 0.2 & 0.14 && 0.0 & \textbf{\color{blue}{0.3}} & 0.16 \\

624 & 0.0 && 0.0 & 0.0 & 0.0 && 0.0 & 0.0 & 0.0 && 0.0 & 0.0 & 0.0 && 0.0 & 0.0 & 0.0 && 0.0 & 0.0 & 0.0 && 0.0 & 0.0 & 0.0 && 0.0 & 0.0 & 0.0 \\

625 & 0.0 && 0.0 & 0.0 & 0.0 && 0.0 & 0.0 & 0.0 && 0.0 & 0.0 & 0.0 && 0.0 & 0.0 & 0.0 && 0.0 & 0.0 & 0.0 && 0.0 & 0.0 & 0.0 && 0.0 & 0.0 & 0.0 \\

626 & 0.0 && 0.0 & 0.0 & 0.0 && 0.0 & 0.0 & 0.0 && 0.0 & 0.0 & 0.0 && 0.0 & 0.0 & 0.0 && 0.0 & 0.0 & 0.0 && 0.0 & 0.0 & 0.0 && 0.0 & 0.0 & 0.0 \\

627 & 0.0 && 0.0 & 0.0 & 0.0 && 0.0 & 0.0 & 0.0 && 0.0 & 0.0 & 0.0 && 0.0 & 0.0 & 0.0 && 0.0 & 0.0 & 0.0 && 0.0 & 0.0 & 0.0 && 0.0 & 0.0 & 0.0 \\

628 & 0.0 && 0.0 & 0.0 & 0.0 && 0.0 & 0.0 & 0.0 && 0.0 & 0.0 & 0.0 && 0.0 & 0.0 & 0.0 && 0.0 & 0.0 & 0.0 && 0.0 & 0.0 & 0.0 && 0.0 & 0.0 & 0.0 \\

629 & 0.7 && 0.0 &  \textbf{\color{blue}{0.6}} &  \textbf{\color{red}{0.4}} && 0.0 & \textbf{\color{blue}{0.6}} & 0.25
 && \textbf{0.1} & 0.4 & 0.24 && \textbf{0.1} & 0.3 & 0.19 && 0.0 & \textbf{\color{blue}{0.6}} & 0.33 && \textbf{0.1} & \textbf{\color{blue}{0.6}} & 0.32 
 && \textbf{0.1} & 0.3 & 0.11 \\
\cmidrule(lr){1-30}
\end{tabular*}}
\end{table*}

\subsubsection{Diversity}
\begin{figure}[!h]
\centering
\includegraphics[width=0.5 \linewidth]{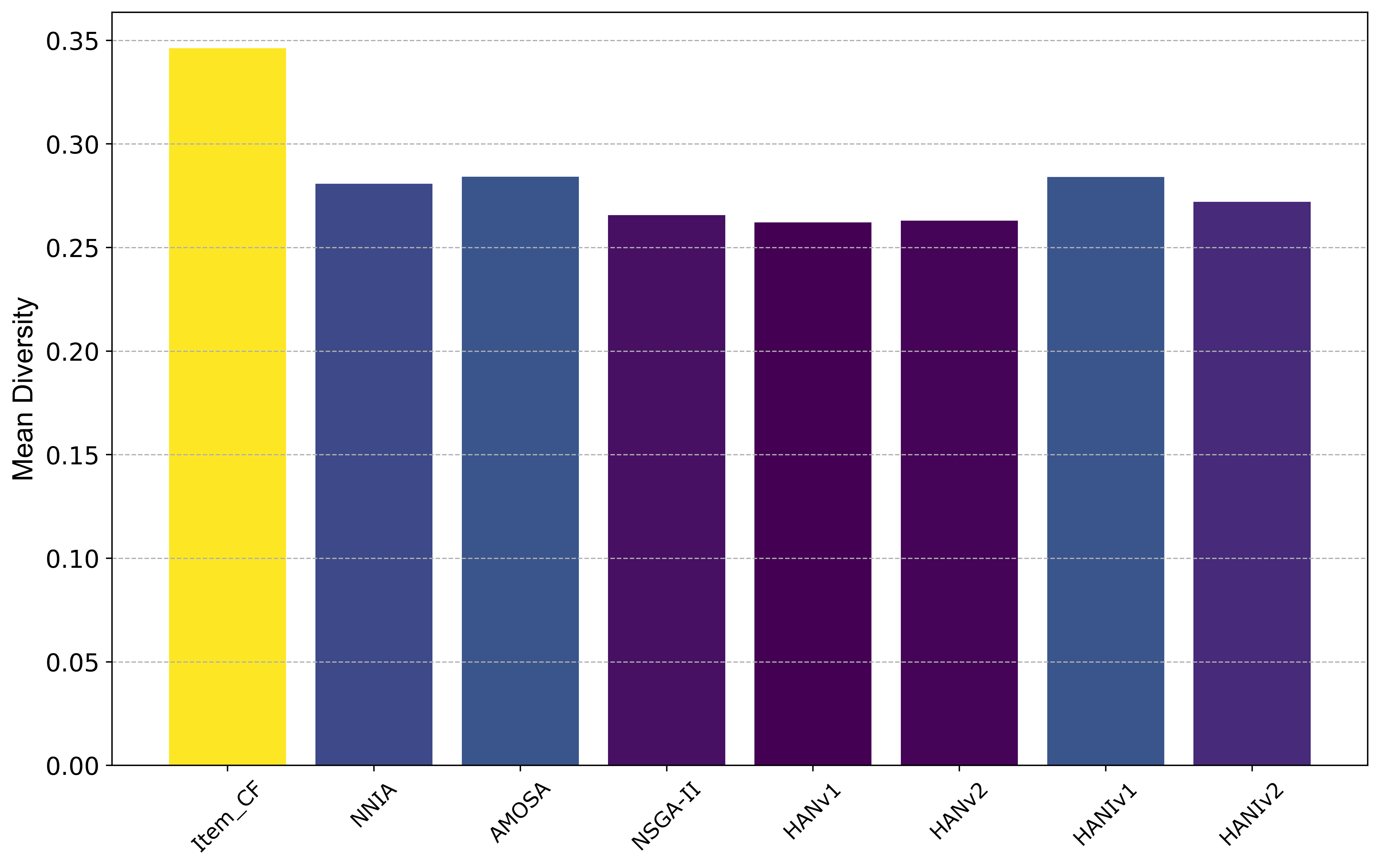}
\caption{Diversity comparisons on MovieLens.}\label{fig:divcommovie}
\end{figure}

Table \ref{tab:Diversity:Movie} presents the diversity results of the recommendation lists on the MovieLens dataset, where lower values indicate more diverse recommendations. In terms of minimum values, NNIA and HANIv1 perform best, while HANv1 achieves the most diverse recommendations in terms of maximum values. Looking at mean values, HANv1, HANv2, and HANIv2 demonstrate strong performance, with HANv1 and HANv2 particularly efficient at generating diverse recommendation lists, as shown in Figure \ref{fig:divcommovie}.

\begin{figure}[!h]
\centering
\includegraphics[width=0.5 \linewidth]{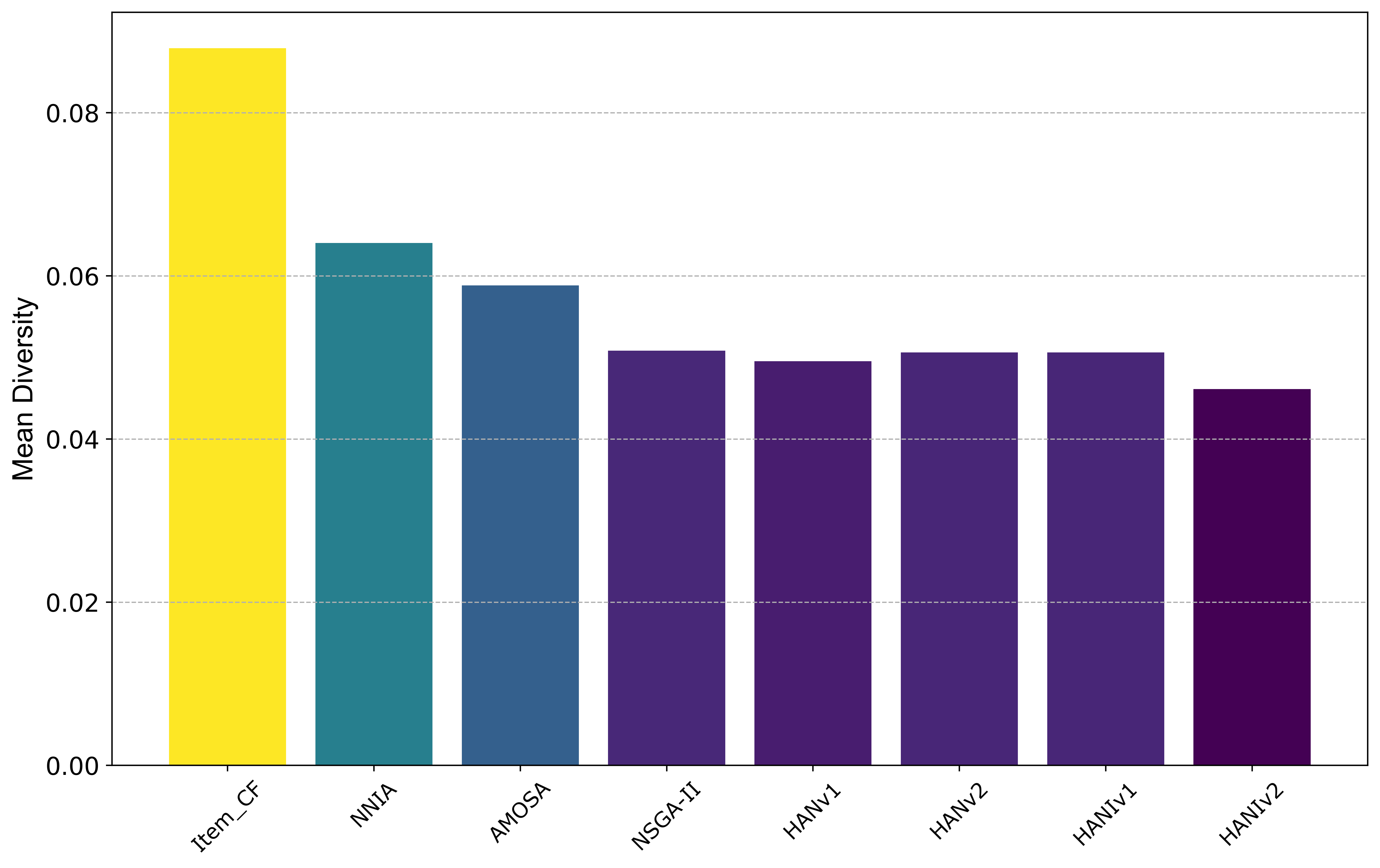}
\caption{Diversity comparisons on ModCloth.}\label{fig:divcomModCloth}
\end{figure}

Table \ref{tab:Diversity:ModCloth} illustrates the diversity outcomes on the ModCloth dataset. Here, NNIA and HANIv1 perform well for minimum diversity values. HANv1 exhibits superior efficiency for maximum values, while HANIv2 shows the best performance in terms of mean diversity values. This strong performance by HANIv2 is further reflected in Figure \ref{fig:divcomModCloth}.

\begin{table*}[!ht] 
\caption{\label{tab:Diversity:Movie} Diversity on MovieLens.}
    \scalebox{0.4}{
    \begin{tabular*}{\linewidth}{@{\extracolsep{\fill}}cccccccccccccccccccccccccccccccc@{}}
\cmidrule(lr){1-30}
      UserID & Item-CF &&& NNIA &  &&&NSGAII&&&  &AMOSA&&&  & HANv1&&&  & HANv2 &&&  & HANIv1 &&&  & HANIv2\\
        \cmidrule(lr){4-6}\cmidrule(lr){8-10}\cmidrule(lr){12-14}\cmidrule(lr){16-18}\cmidrule(lr){20-22}\cmidrule(lr){24-26}\cmidrule(lr){28-30}
          &&&$\min$ & $\max$ & mean&& $\min$& $\max$& mean&& $\min$& $\max$& mean&& $\min$& $\max$& mean&& $\min$& $\max$& mean&& $\min$& $\max$& mean&& $\min$& $\max$& mean\\
\cmidrule(lr){1-30}
3411 & 0.3723 && \textbf{0.18} & 0.4 & 0.31 && 0.2 & 0.38 & 0.28 && 0.19 & 0.39 & 0.3 && 0.22 &  \textbf{\color{blue}{0.33}} & \textbf{\color{red}{0.27}} 
&& 0.19 & 0.39 & 0.28 &&\textbf{0.18} & 0.4 & 0.3 && 0.2 & 0.39 & 0.31 \\

3412 & 0.4012 && \textbf{0.18} & 0.42 & 0.31 && 0.2 & 0.39 & 0.29 && 0.19 & 0.39 & 0.29 && 0.22 &  \textbf{\color{blue}{0.34}} &  \textbf{\color{red}{0.28}} && 0.19 & 0.4 &  \textbf{\color{red}{0.28}} && \textbf{0.18} & 0.42 & 0.3 && 0.2 & 0.39 & 0.3 \\

3413 & 0.3297 && \textbf{0.14} & 0.41 & 0.28 && 0.15 & 0.39 & 0.26 && 0.15 & 0.38 & 0.27 && 0.19 &  \textbf{\color{blue}{0.33}} &  \textbf{\color{red}{0.25}} &&\textbf{0.14} & 0.4 & 0.26 &&\textbf{0.14} & 0.42 & 0.27 && 0.15 & 0.38 &  \textbf{\color{red}{0.25}} \\

3414 & 0.3635 && \textbf{0.16} & 0.39 & 0.28 && 0.18 & 0.38 & 0.27 && 0.17 & 0.37 & 0.28 && 0.2 &  \textbf{\color{blue}{0.34}} & 0.26 && \textbf{0.16} & 0.37 & 0.26 && \textbf{0.16} & 0.39 & 0.28 && 0.17 & 0.37 &  \textbf{\color{red}{0.24}} \\

3415 & 0.3449 && \textbf{0.17} & 0.43 & 0.29 && 0.18 & 0.41 & 0.28 && 0.18 & 0.42 & 0.3 && 0.2 &  \textbf{\color{blue}{0.36}} & 0.28 && \textbf{0.17} & 0.42 & 0.28 && \textbf{0.17} & 0.44 & 0.3 && 0.18 & 0.42 &  \textbf{\color{red}{0.26}} \\

3416 & 0.3263 &&  \textbf{0.14} & 0.43 & \textbf{\color{red}{ 0.25}} && 0.16 & 0.37 & \textbf{\color{red}{ 0.25}}  && 0.16 & 0.4 & 0.27 && 0.19 &  \textbf{\color{blue}{0.33}} & \textbf{\color{red}{ 0.25}}  && 0.15 & 0.4 & 0.26 && 0.15 & 0.43 & 0.32 && 0.16 & 0.4 & 0.29 \\

3417 & 0.2844 && \textbf{0.11} & 0.36 & 0.27 && 0.12 & 0.35 & \textbf{\color{red}{0.22}} && 0.12 & 0.34 & 0.24
 && 0.16 &  \textbf{\color{blue}{0.31}} & 0.23 && 0.12 & 0.35 & \textbf{\color{red}{0.22}} && \textbf{0.11} & 0.36 & 0.26 && 0.13 & 0.34 & 0.27 \\

3418 & 0.3285 && \textbf{0.13} & 0.38 & \textbf{\color{red}{0.24}} && 0.16 & 0.37 & 0.25 && 0.15 & 0.36 & 0.28 && 0.17 &  \textbf{\color{blue}{0.32}} & 0.25 && 0.14 & 0.37 & \textbf{\color{red}{0.24}} && \textbf{0.13} & 0.38 & 0.26 && 0.15 & 0.36 & 0.26 \\

3419 & 0.371 && \textbf{0.16} & 0.47 & 0.31 && 0.18 & 0.42 & 0.29 && 0.18 & 0.45 & 0.33 && 0.22 &  \textbf{\color{blue}{0.37}} & 0.29 && 0.17 & 0.45 & 0.29 && \textbf{0.16} & 0.47 & 0.27 && 0.18 & 0.45 & \textbf{\color{red}{0.25}} \\

3420 & 0.3402 && \textbf{0.17} & 0.39 & 0.29 && 0.18 & 0.37 & 0.27 && 0.18 & 0.37 & 0.28 
&& 0.2 &  \textbf{\color{blue}{0.33}} & \textbf{\color{red}{0.26}} && 0.18 & 0.37 & \textbf{\color{red}{0.26}} && \textbf{0.17} & 0.39 & 0.29 && 0.19 & 0.38 & 0.28 \\
\cmidrule(lr){1-30}
    \end{tabular*}}
\end{table*}

\begin{table*}[!ht] 
\caption{\label{tab:Diversity:ModCloth} Diversity on  ModCloth.}
    \scalebox{0.4}{
    \begin{tabular*}{\linewidth}{@{\extracolsep{\fill}}cccccccccccccccccccccccccccccccc@{}}
\cmidrule(lr){1-30}
      UserID & Item-CF &&& NNIA &  &&&NSGAII&&&  &AMOSA&&&  & HANv1&&&  & HANv2 &&&  & HANIv1 &&&  & HANIv2\\
        \cmidrule(lr){4-6}\cmidrule(lr){8-10}\cmidrule(lr){12-14}\cmidrule(lr){16-18}\cmidrule(lr){20-22}\cmidrule(lr){24-26}\cmidrule(lr){28-30}
          &&&$\min$ & $\max$ & mean&& $\min$& $\max$& mean&& $\min$& $\max$& mean&& $\min$& $\max$& mean&& $\min$& $\max$& mean&& $\min$& $\max$& mean&& $\min$& $\max$& mean\\
\cmidrule(lr){1-30}
620 & 0.0876 && \textbf{0.03} & 0.1 & 0.06 && 0.04 & 0.09 & 0.06 && 0.04 & 0.09 & 0.06 && 0.04 &  \textbf{\color{blue}{0.08}} & 0.06 && 0.04 & 0.09 & 0.06 &&
\textbf{ 0.03} & 0.1 & 0.07 && 0.04 & 0.09 & \textbf{\color{red}{0.05}}\\

621 & 0.0911 && \textbf{0.01} & 0.09 & 0.05 && 0.02 & 0.09 & 0.05 && 0.04 & 0.08 & 0.06 && 0.03 &  \textbf{\color{blue}{0.07}} &\textbf{\color{red}{ 0.04}} && 0.02 & 0.09 & 0.05 && \textbf{0.01} & 0.09 &\textbf{\color{red}{ 0.04}} && 0.03 & 0.08 & 0.06 \\ 

622 & 0.0791 && \textbf{0.01} & 0.1 & \textbf{\color{red}{0.05}} && 0.02 & 0.1 & \textbf{\color{red}{0.05}}
 && 0.04 &  \textbf{\color{blue}{0.08}} & 0.07 && 0.03 & 0.08 & \textbf{\color{red}{0.05}} && 
 0.02 &  \textbf{\color{blue}{0.08}} &\textbf{\color{red}{0.05}} && \textbf{0.01} & 0.1 & 0.06 && 0.06 &  \textbf{\color{blue}{0.08}} & 0.08 \\ 

623 & 0.0895 && \textbf{0.03} & 0.09 & 0.07 && 0.04 & 0.09 & \textbf{\color{red}{0.06}} && 0.04 & 0.09 & \textbf{\color{red}{0.06}}
 && 0.04 & \textbf{\color{blue}{0.08}} & \textbf{\color{red}{0.06}} &&\textbf{0.03} & 0.09 & \textbf{\color{red}{0.06}}
  && \textbf{0.03} & 0.1 & \textbf{\color{red}{0.06}} && 0.05 & 0.09 & 0.07 \\ 

624 & 0.0936 && \textbf{0.01} & 0.1 & 0.06 && 0.02 & 0.09 & 0.05 && 0.03 & \textbf{\color{blue}{0.07}} & 0.05 && 0.03 & 0.08 & 0.05 && 0.02 & 0.08 & 0.05 && \textbf{0.01} & 0.1 & 0.05 && 0.03 & 0.09 & \textbf{\color{red}{0.04}} \\ 

625 & 0.0911 && \textbf{0.01} & 0.1 & 0.06 && \textbf{0.01} &\textbf{\color{blue}{0.08}} & \textbf{\color{red}{0.04}}
 && 0.03 & \textbf{\color{blue}{0.08}} & 0.06 && 0.03 &\textbf{\color{blue}{0.08}} & 0.05 && 0.02 & 0.09 & 0.05 &&\textbf{0.01} & 0.1 & \textbf{\color{red}{0.04}}
  && 0.03 &\textbf{\color{blue}{0.08}} & \textbf{\color{red}{0.04}} \\ 

626 & 0.0936 && \textbf{0.01} & 0.1 & 0.09 && 0.02 & 0.09 & 0.05 && 0.03 & 0.08 & 0.05 && 0.02 &\textbf{\color{blue}{0.07}} & 0.04 && \textbf{0.01} & 0.08 & 0.04 &&\textbf{0.01} & 0.09 & 0.04 && 0.03 & 0.08 &\textbf{\color{red}{0.03}}\\

627 & 0.089 && 0.02 & 0.1 & 0.07 && 0.02 & 0.09 & 0.05 && 0.03 &\textbf{\color{blue}{0.08}} & 0.07 && 0.03 & \textbf{\color{blue}{0.08}}&
 0.06 && 0.02 & 0.09 & 0.05 && \textbf{0.01} & 0.09 & 0.05 && 0.03 & \textbf{\color{blue}{0.08}} &  \textbf{\color{red}{0.03}}\\

628 & 0.0874 && \textbf{0.02 }& 0.1 & 0.05 && 0.03 & 0.09 & 0.06 && 0.04 & \textbf{\color{blue}{0.08}} & 0.06 && 0.03 & \textbf{\color{blue}{0.08}}
 & 0.05 && 0.03 & 0.1 & 0.06 && \textbf{0.02} & 0.1 & 0.05 && 0.03 & \textbf{\color{blue}{0.08}} & \textbf{\color{red}{ 0.04}}\\

629 & 0.0771 && \textbf{0.01} & 0.09 & 0.07 && 0.02 & 0.08 & 0.04 && 0.02 & 0.06 & 0.05 && 0.02 & \textbf{\color{blue}{0.04}} & \textbf{\color{red}{ 0.03}} && 0.02 & 0.08 & 0.05 &&
\textbf{ 0.01} & 0.09 & 0.05 && 0.02 & 0.06 &\textbf{\color{red}{ 0.03}} \\ 
\cmidrule(lr){1-30}
    \end{tabular*}}
\end{table*}

\subsubsection{Novelty}
\begin{figure}[!h]
\centering
\includegraphics[width=0.5 \linewidth]{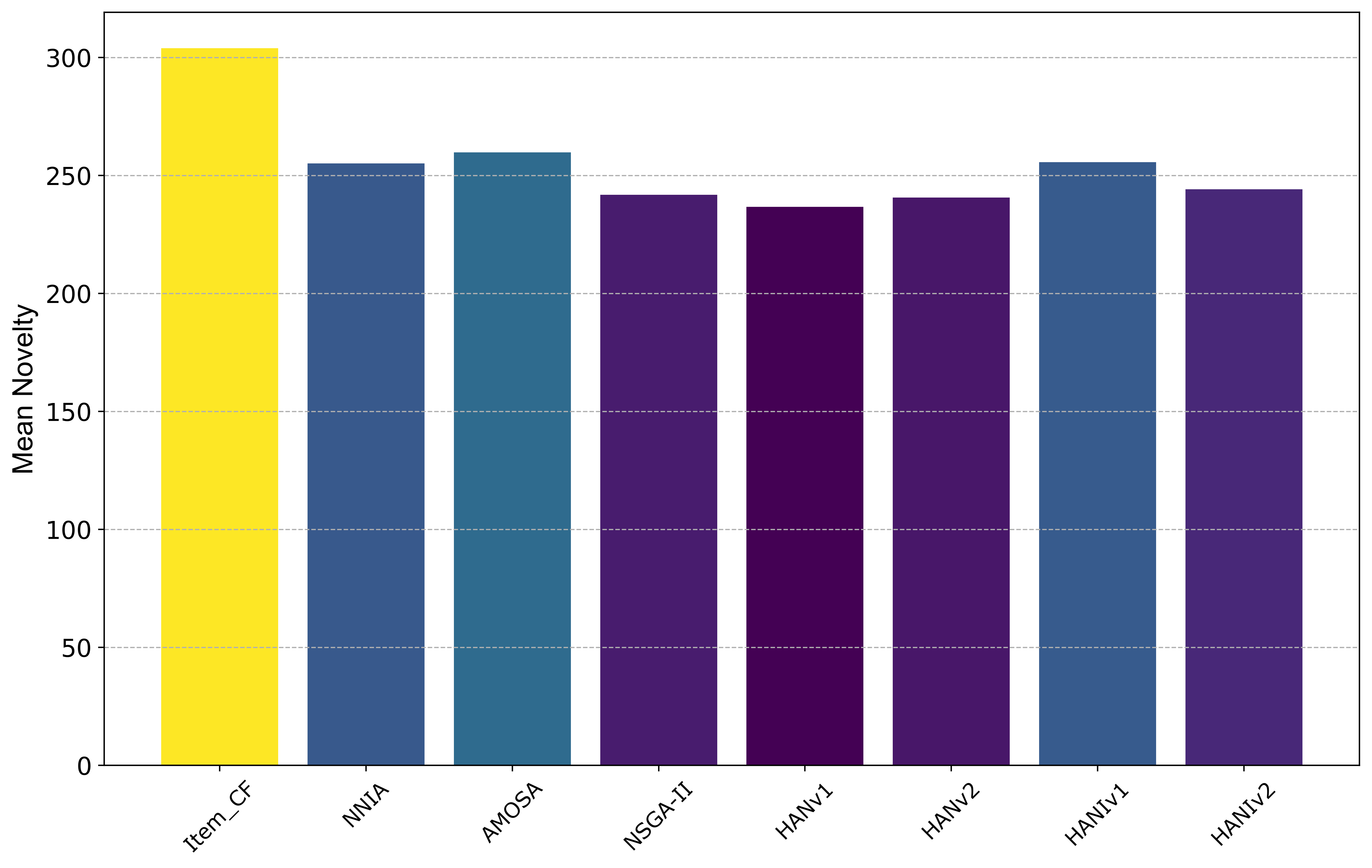}
\caption{Novelty comparisons on MovieLens.}\label{fig:novcommovie}
\end{figure}

Table \ref{tab:Novelty:Movie} presents the novelty results of the recommendation lists on the MovieLens dataset, where lower values indicate more novel recommendations. In terms of minimum values, NNIA and HANIv1 achieve better novelty outcomes, while HANv1 excels with the best results for maximum values. Regarding mean values, both HANv1 and HANIv2 demonstrate stronger performance, indicating their ability to consistently generate more novel recommendations. Figure \ref{fig:novcommovie} further illustrates the mean novelty results across users, highlighting the efficiency of HANv1 in delivering novel recommendation lists.

\begin{figure}[!h]
\centering
\includegraphics[width=0.5 \linewidth]{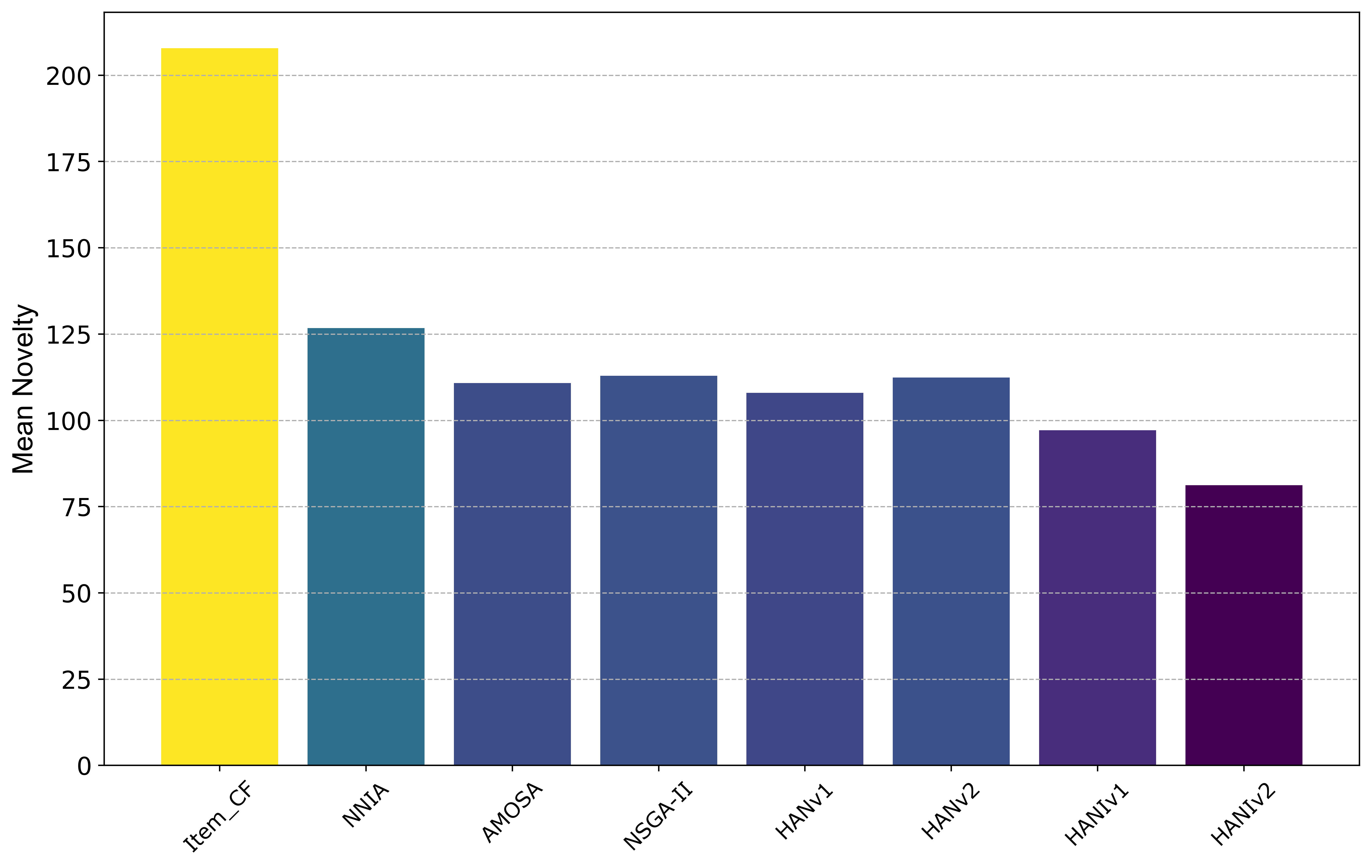}
\caption{Novelty comparisons on ModCloth.}\label{fig:novcomModCloth}
\end{figure}

In Table \ref{tab:Novelty:ModCloth}, the novelty results for the ModCloth dataset are presented. NNIA shows the best performance in terms of minimum values, while HANv1 produces the smallest maximum values. HANIv2 stands out for achieving the smallest mean values, demonstrating its effectiveness in generating novel recommendations. Figure \ref{fig:novcomModCloth} reinforces this performance.

 \begin{table*}[!h]
\caption{\label{tab:Novelty:Movie} Novelty on MovieLens.}
   \scalebox{0.35}{
    \begin{tabular*}{\linewidth}{@{\extracolsep{\fill}}cccccccccccccccccccccccccccccccc@{}}
\cmidrule(lr){1-30}
UserID & Item-CF &&& NNIA &  &&&NSGAII&&&  &AMOSA&&&  & HANv1&&&  & HANv2 &&&  & HANIv1 &&&  & HANIv2\\
\cmidrule(lr){4-6}\cmidrule(lr){8-10}\cmidrule(lr){12-14}\cmidrule(lr){16-18}\cmidrule(lr){20-22}\cmidrule(lr){24-26}\cmidrule(lr){28-30}
          &&&$\min$ & $\max$ & mean&& $\min$& $\max$& mean&& $\min$& $\max$& mean&& $\min$& $\max$& mean&& $\min$& $\max$& mean&& $\min$& $\max$& mean&& $\min$& $\max$& mean\\
\cmidrule(lr){1-30}
3411 & 364.8 && 174.62 & 388.16 & 297.46 && 184.58 & 351.74 & 255.73 && 178.18 & 369.12 & 278.93 && 192.64 &  \textbf{\color{blue}{296.84}} &  \textbf{\color{red}{241.91}} 
&& 181.4 & 362.12 & 256.03 && \textbf{171.96} & 376.6 & 275.27 && 186.58 & 371.2 & 288.67 \\ 

3412 & 394.9 && \textbf{158.74} & 427.1 & 298.24 && 184.66 & 375.9 & 269.49 && 164.82 & 384.42 & 267.87 && 196.54 & \textbf{\color{blue}{328.08}} & \textbf{\color{red}{256.17}} 
&& 163.34 & 402.6 & 268.34 && 161.76 & 428.72 & 292.76 && 174.92 & 387.18 & 276.16 \\

3413 & 288.0 && \textbf{121.48} & 424.34 & 283.33 && 137.08 & 403.24 & 256.8 && 129.4 & 391.68 & 267.15 && 173.66 &  \textbf{\color{blue}{334.22}} & 253.41 && 126.9 & 410.82 & 256.88 && 127.5 & 429.08 & 267.73 && 129.98 & 387.6 & \textbf{\color{red}{247.66}} \\

3414 & 293.3 && 123.36 & 349.12 & 240.18 && 136.56 & 341.04 & 232.51 && 125.24 & 344.96 & 239.47 && 153.52 & \textbf{\color{blue}{297.36}} & 216.41 && \textbf{122.24} & 328.74 & 225.45 && 126.32 & 347.36 & 239.54 && 125.22 & 339.26 & \textbf{\color{red}{202.0}} \\

3415 & 312.4 && \textbf{124.36} & 424.54 & 255.44 && 148.78 & 402.64 & 266.64 && 134.74 & 442.46 & 285.63 && 169.78 & \textbf{\color{blue}{351.56}} & 265.5 && 134.16 & 422.38 & 265.95 && 125.76 & 451.46 & 282.02 && 145.84 & 438.48 &  \textbf{\color{red}{242.59}} \\

3416 & 299.7 && 114.22 & 342.46 & 202.59 && \textbf{112.84} & 303.8 &  \textbf{\color{red}{194.07}} && 116.7 & 332.7 & 232.64 && 140.22 & \textbf{\color{blue}{278.28}} & 197.43 && 114.44 & 312.68 & 201.7 && 115.34 & 333.18 & 247.08 && 126.46 & 329.1 & 244.49 \\

3417 & 93.4 && 57.64 & 265.44 & 187.4 && 77.18 & 267.7 & 173.6 && \textbf{56.5} & 253.48 & 174.03 && 108.22 & \textbf{\color{blue}{234.14}} & 176.26 && 61.58 & 259.32 & \textbf{\color{red}{161.63}} && 58.22 & 258.6 & 177.99 && 78.0 & 251.86 & 194.03 \\

3418 & 303.4 && 109.86 & 418.92 & \textbf{\color{red}{238.52}} && 147.02 & 405.18 & 253.93 && 122.3 & 388.5 & 276.48 && 174.72 & \textbf{\color{blue}{319.12}} & 247.0 && 120.0 & 407.32 & 244.91 && \textbf{108.2} & 416.6 & 257.93 && 129.28 & 391.24 & 260.8 \\

3419 & 361.0 && 133.06 & 451.02 & 280.51 && 152.84 & 405.3 & 269.52 && 143.64 & 453.92 & 327.82 && 180.32 &  \textbf{\color{blue}{362.74}} & 276.63 && 141.52 & 449.9 & 281.81 && \textbf{127.34} & 448.84 & 249.94 && 152.24 & 449.82 & \textbf{\color{red}{236.64}} \\

3420 & 328.0 && 130.82 & 381.62 & 267.64 && 140.92 & 371.92 & 245.91 && 145.18 & 359.66 & 248.11 && 167.84 &  \textbf{\color{blue}{313.44}} & \textbf{\color{red}{235.76}} && 137.16 & 355.54 & 243.45 && \textbf{128.94} & 388.22 & 265.76 && 148.6 & 354.84 & 248.54 \\
\cmidrule(lr){1-30}
\end{tabular*}}
\end{table*}

 \begin{table*}[!h]
\caption{\label{tab:Novelty:ModCloth} Novelty on ModCloth.}
   \scalebox{0.4}{
    \begin{tabular*}{\linewidth}{@{\extracolsep{\fill}}cccccccccccccccccccccccccccccccc@{}}
\cmidrule(lr){1-30}
UserID & Item-CF &&& NNIA &  &&&NSGAII&&&  &AMOSA&&&  & HANv1&&&  & HANv2 &&&  & HANIv1 &&&  & HANIv2\\
\cmidrule(lr){4-6}\cmidrule(lr){8-10}\cmidrule(lr){12-14}\cmidrule(lr){16-18}\cmidrule(lr){20-22}\cmidrule(lr){24-26}\cmidrule(lr){28-30}
          &&&$\min$ & $\max$ & mean&& $\min$& $\max$& mean&& $\min$& $\max$& mean&& $\min$& $\max$& mean&& $\min$& $\max$& mean&& $\min$& $\max$& mean&& $\min$& $\max$& mean\\
\cmidrule(lr){1-30}\\[-0.3cm]
620 & 202.6 && 45.0 & 146.7 & 89.9 && 47.2 & 147.4 & 100.41 && 51.1 & 152.5 & 118.34 && 70.9 & 169.1 & 111.37 && 59.3 & 159.2 & 104.99 && \textbf{40.3} &  \textbf{\color{blue}{143.9}} & 92.1 && 51.1 & 148.1 & \textbf{\color{red}{82.89}} \\

621 & 202.7 &&\textbf{ 21.3} & 197.3 & 105.33 && 57.0 & 206.3 & 120.75 && 65.5 & 151.9 & 110.63 && 65.0 & \textbf{\color{blue}{148.8}} & 97.49 && 27.9 & 209.8 & 113.17 && 25.8 & 186.7 &\textbf{\color{red}{72.46}} && 47.5 & 183.9 & 117.48 \\

622 & 170.8 && \textbf{24.0} & 175.0 & \textbf{\color{red}{93.4}} && 32.2 & 191.6 & 117.45 && 80.7 & \textbf{\color{blue}{149.4}} & 116.75 && 58.3 & 162.6 & 110.01 && 33.0 & 163.0 & 98.72 && 28.9 & 199.8 & 121.19 && 108.2 & 154.9 & 136.98 \\

623 & 190.2 && 45.3 & 145.8 & 106.6 && 76.9 & 169.3 & 111.43 && 59.7 & 156.2 & 97.23 && 71.5 &  \textbf{\color{blue}{141.9}} & 113.88 && 35.4 & 161.3 & 106.92 && \textbf{33.1} & 148.3 & \textbf{\color{red}{79.56}} && 69.6 & 145.0 & 113.1 \\

624 & 250.5 && \textbf{21.0} & 222.0 & 139.89 && 27.1 & 242.7 & 136.34 && 59.3 &  \textbf{\color{blue}{163.7}} & 107.89 && 93.7 & 230.7 & 158.63 && 58.6 & 210.3 & 130.77 && 21.3 & 236.7 & 126.85 && 62.4 & 198.6 & \textbf{\color{red}{74.71}} \\

625 & 202.7 && 27.8 & 206.0 & 134.07 && 21.2 & 177.2 & 98.01 && 47.0 & 150.5 & 101.9 && 65.4 & 153.5 & 113.81 && 55.3 & 188.7 & 112.9 && \textbf{20.5} & 210.8 & 84.02 && 47.0 &  \textbf{\color{blue}{148.1}} & \textbf{\color{red}{58.86}} \\

626 & 250.5 && \textbf{21.1} & 228.6 & 205.65 && 34.4 & 245.1 & 135.34 && 39.3 & 182.9 & 115.74 && 41.3 &  \textbf{\color{blue}{171.5}} & 103.15 && 24.6 & 231.4 & 124.69 && 21.7 & 224.3 & 94.64 && 39.3 & 182.9 & \textbf{\color{red}{51.16}} \\

627 & 242.8 && 27.1 & 234.8 & 170.22 &&\textbf{ 15.9} & 205.3 & 108.52 && 51.8 & 216.4 & 149.85 && 57.5 & \textbf{\color{blue}{182.1}} & 114.51 && 29.3 & 222.5 & 126.21 && 27.7 & 216.2 & 116.87 && 51.8 & 207.2 & \textbf{\color{red}{59.96}} \\

628 & 221.7 && \textbf{33.7} & 206.0 & 101.53 && 48.7 & 222.3 & 127.56 && 60.9 & 162.8 & 116.24 && 78.9 &  \textbf{\color{blue}{157.4}} & 112.12 && 47.2 & 192.2 & 125.66 && 34.3 & 204.5 & 99.89 && 53.6 & 160.0 & \textbf{\color{red}{72.38}} \\

629 & 143.7 && \textbf{17.6} & 156.8 & 120.86 && 25.5 & 135.5 & 73.48 && 40.8 & 92.8 & 73.09 && 33.3 &  \textbf{\color{blue}{65.3}} & 44.99 && 26.8 & 139.1 & 80.22 && 22.0 & 140.1 & 83.85 && 40.8 & 92.8 & \textbf{\color{red}{44.59}} \\
\cmidrule(lr){1-30}
\end{tabular*}}
\end{table*}

\subsubsection{Quality of Pareto frontiers}
Now, we compare multi-objective algorithms based on the metrics introduced for evaluating the quality of Pareto frontiers. Table \ref{tab:smmid:MovieLens}
presents the values for SM and MID, while Table \ref{tab:dmsns:MovieLens} shows DM and SNS for the MovieLens dataset across various users. Additionally, Table \ref{tab:clo:MovieLens} demonstrates the value of CLO and ranks each algorithm based on this value. As a reminder, the algorithm with the highest CLO value is considered the best.

\begin{figure*}[!h]
        {\includegraphics[width=4cm]{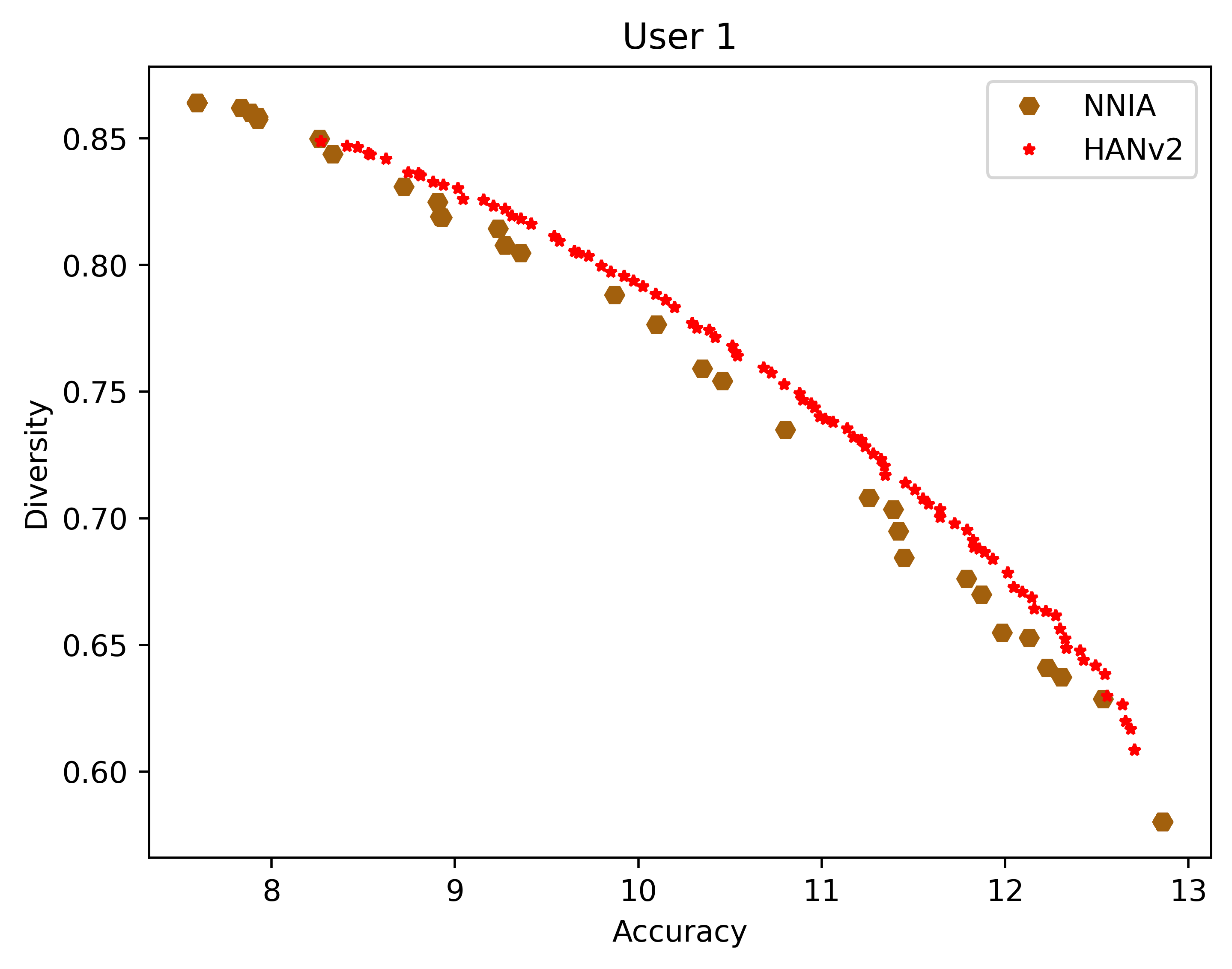}}\quad
        {\includegraphics[width= 4cm]{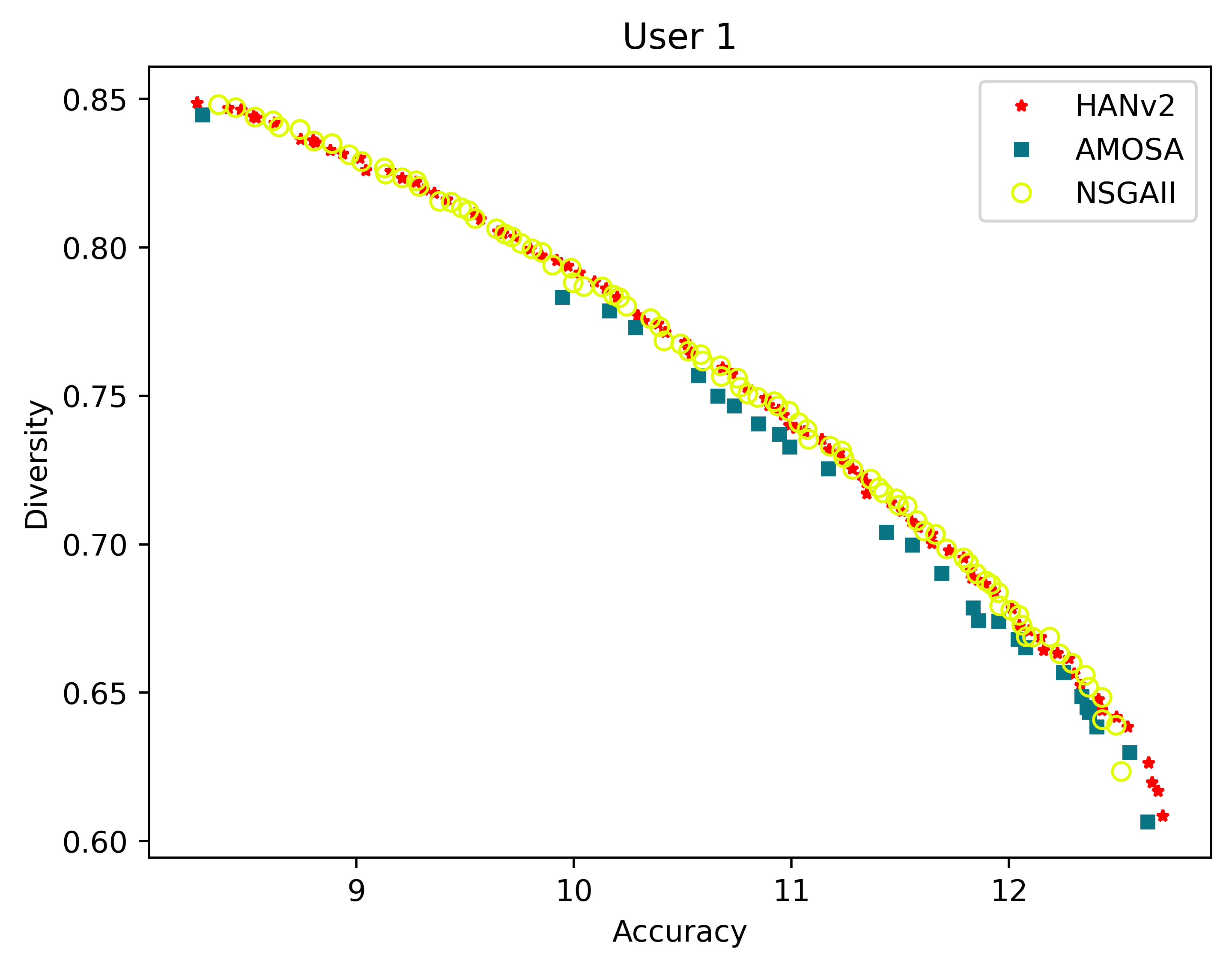}}\quad
        {\includegraphics[width= 4cm]{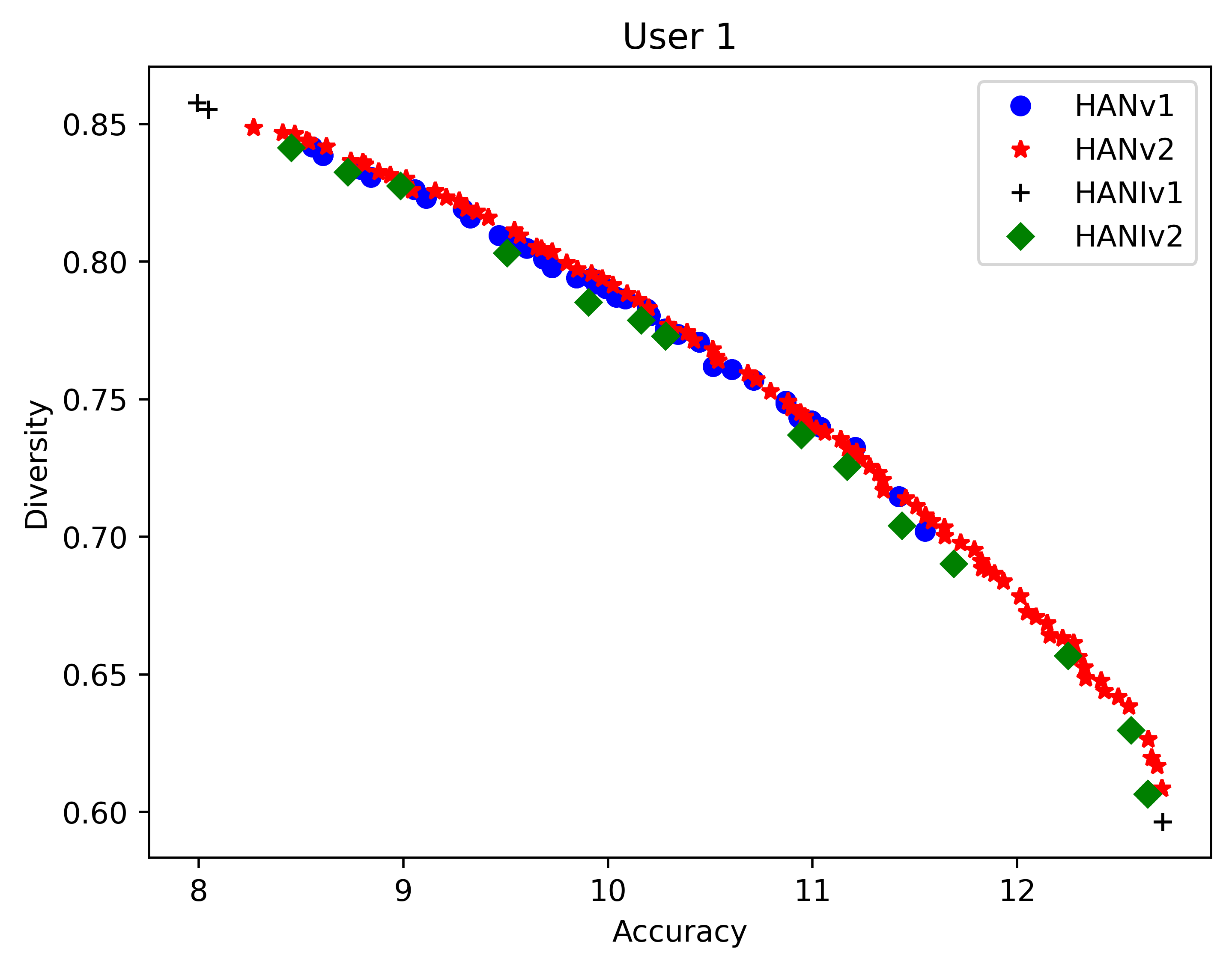}}
        \caption{
        Pareto frontier comparisons for user 1 (MovieLens).}
         \label{fig:paracom:movu1}
    \end{figure*}

\begin{figure*}[!h]
        {\includegraphics[width=4cm]{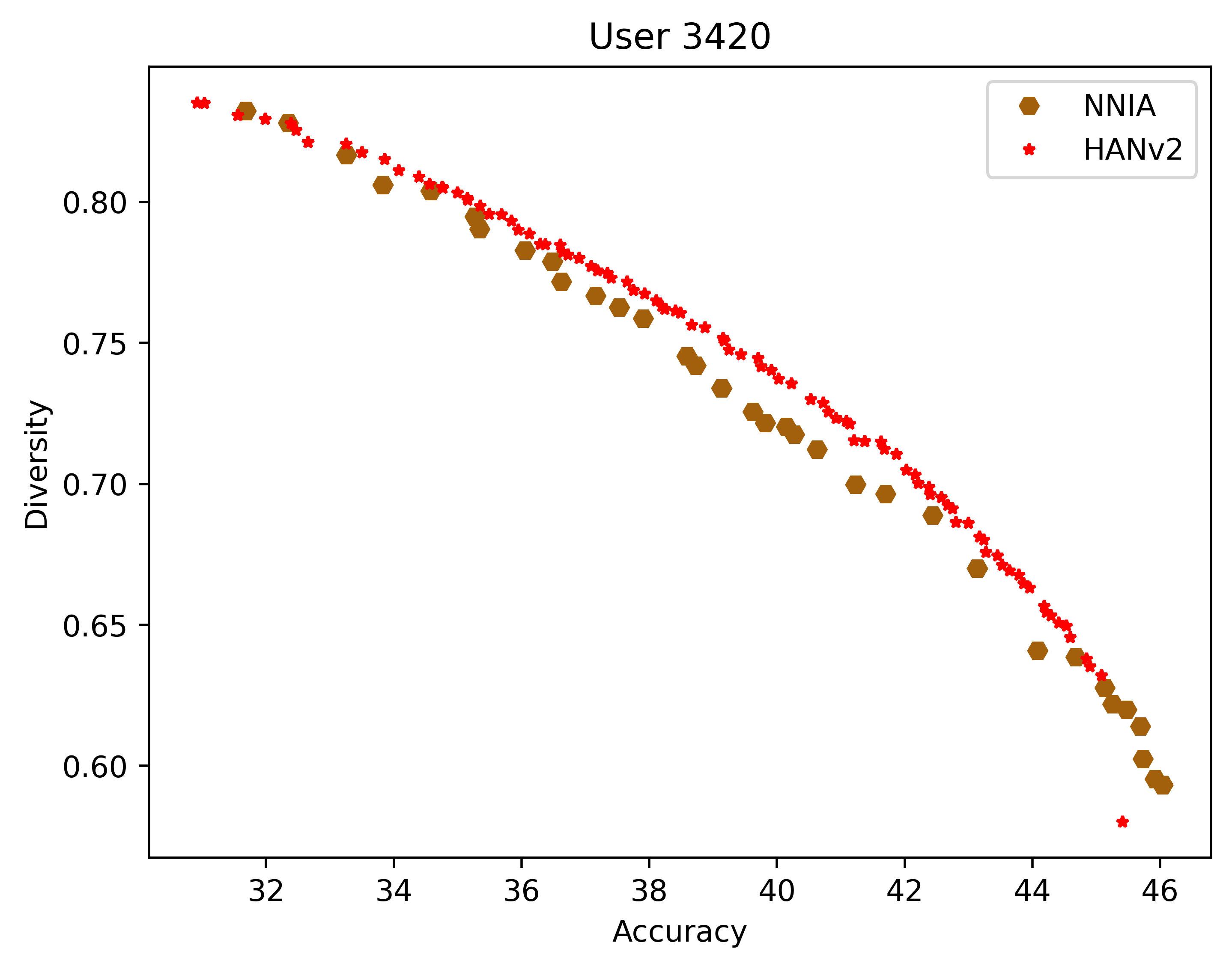}}\quad
        {\includegraphics[width= 4cm]{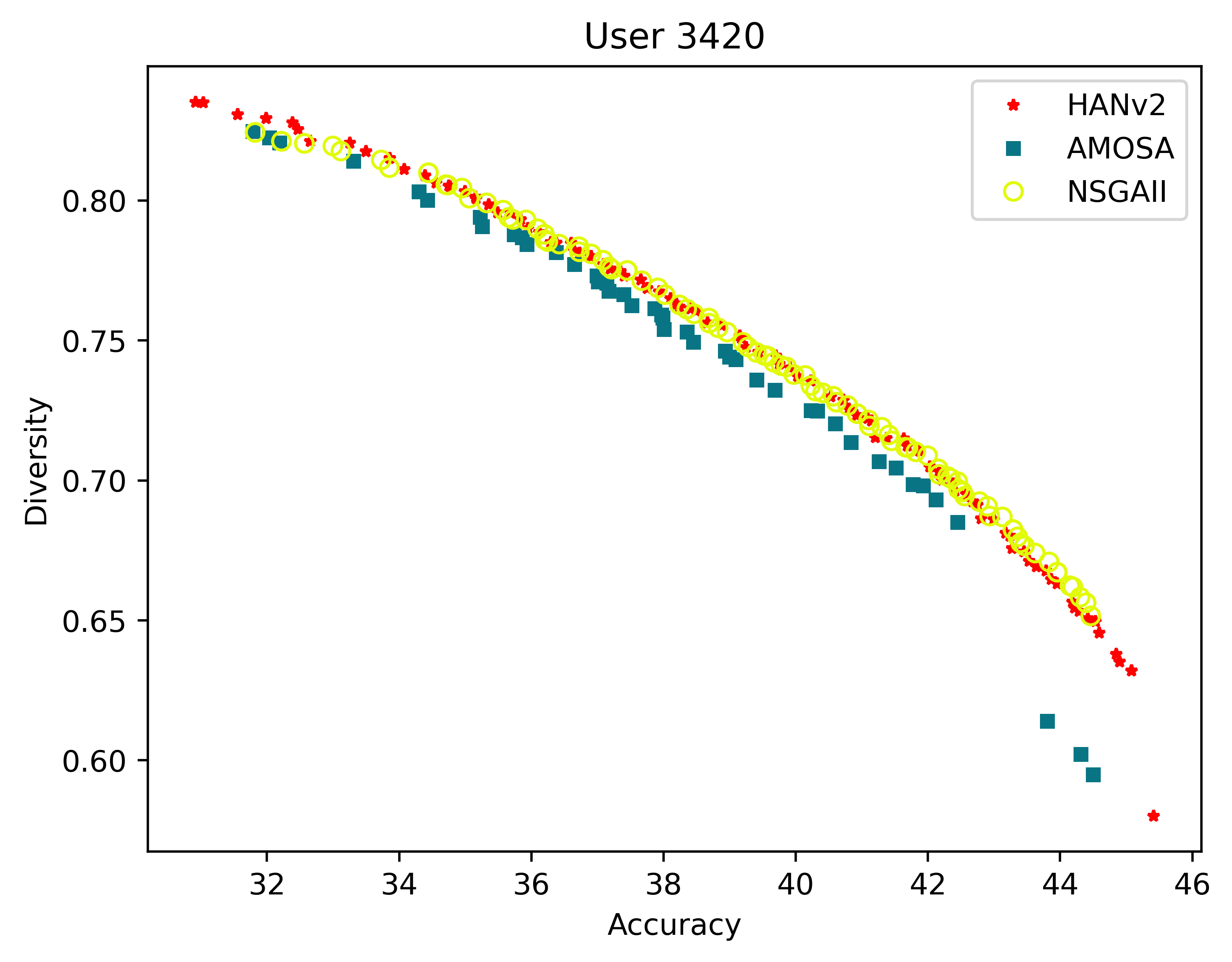}}\quad
        {\includegraphics[width= 4cm]{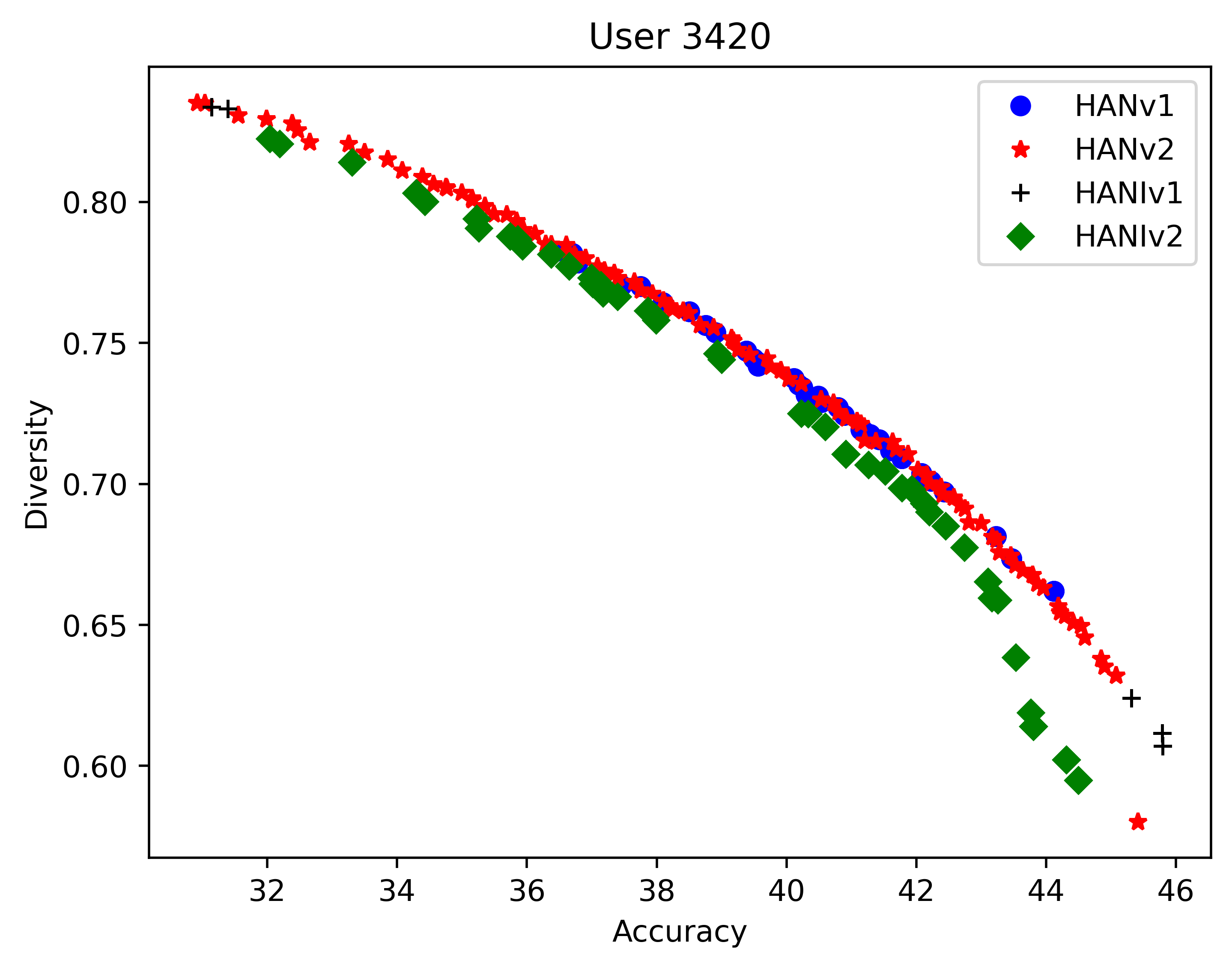}}
        \caption{
        Pareto frontier comparisons for user 3420 (MovieLens).}
         \label{fig:paracom:movu3420}
    \end{figure*}

From these results, it is evident that HANv2 outperforms the other proposed algorithms in creating multi-objective recommender systems based on the four metrics. This superior performance is further illustrated in Figures \ref{fig:paracom:movu1} and \ref{fig:paracom:movu3420}, which show that HANv2 not only outperforms NNIA but also dominates the frontier established by AMOSA. Furthermore, HANv2 generates a more extended frontier than NSGA-II, demonstrating better diversity and uniformity compared to other hybrid algorithms.

\begin{figure*}[!h]
        {\includegraphics[width=4cm]{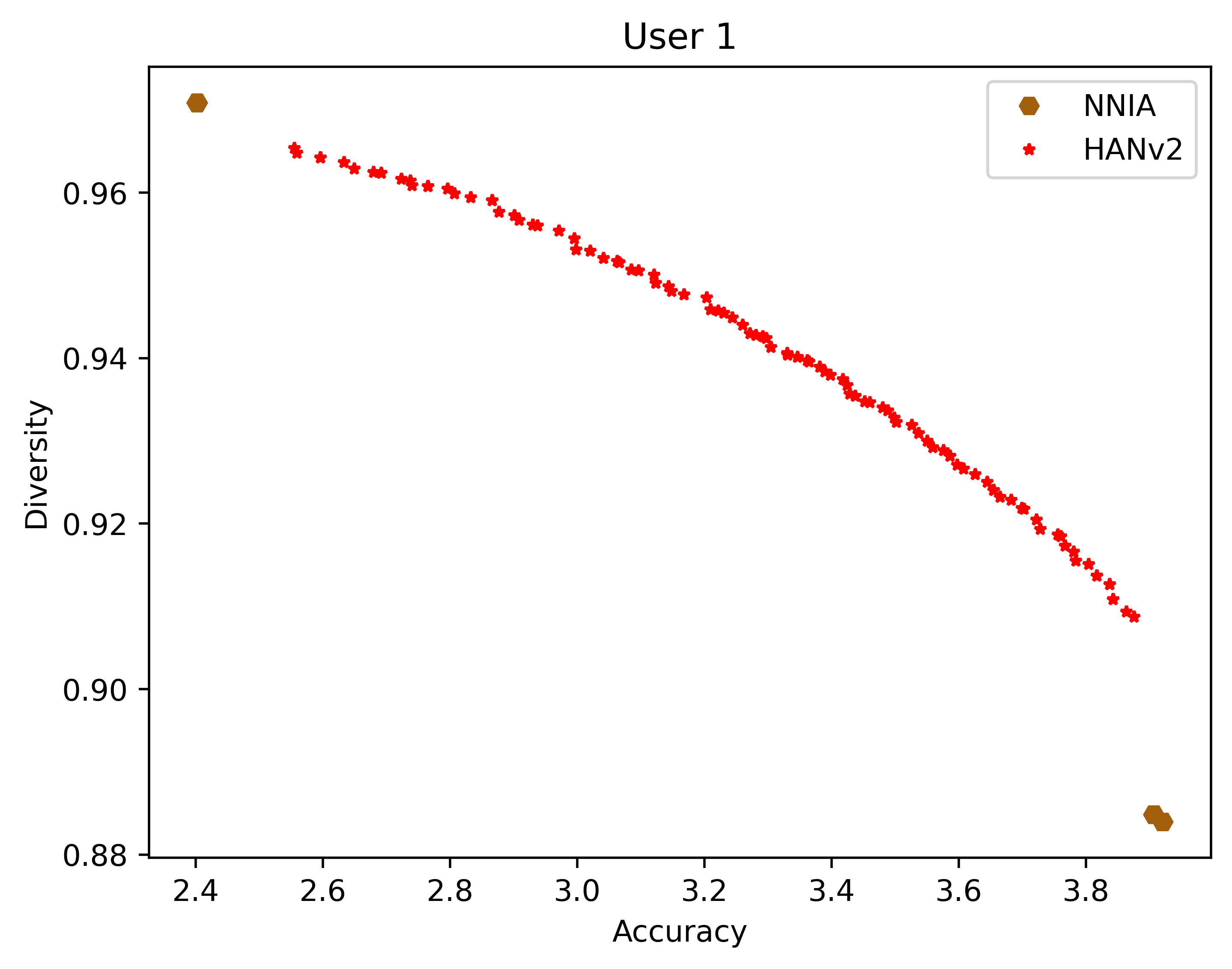}}\quad
        {\includegraphics[width= 4cm]{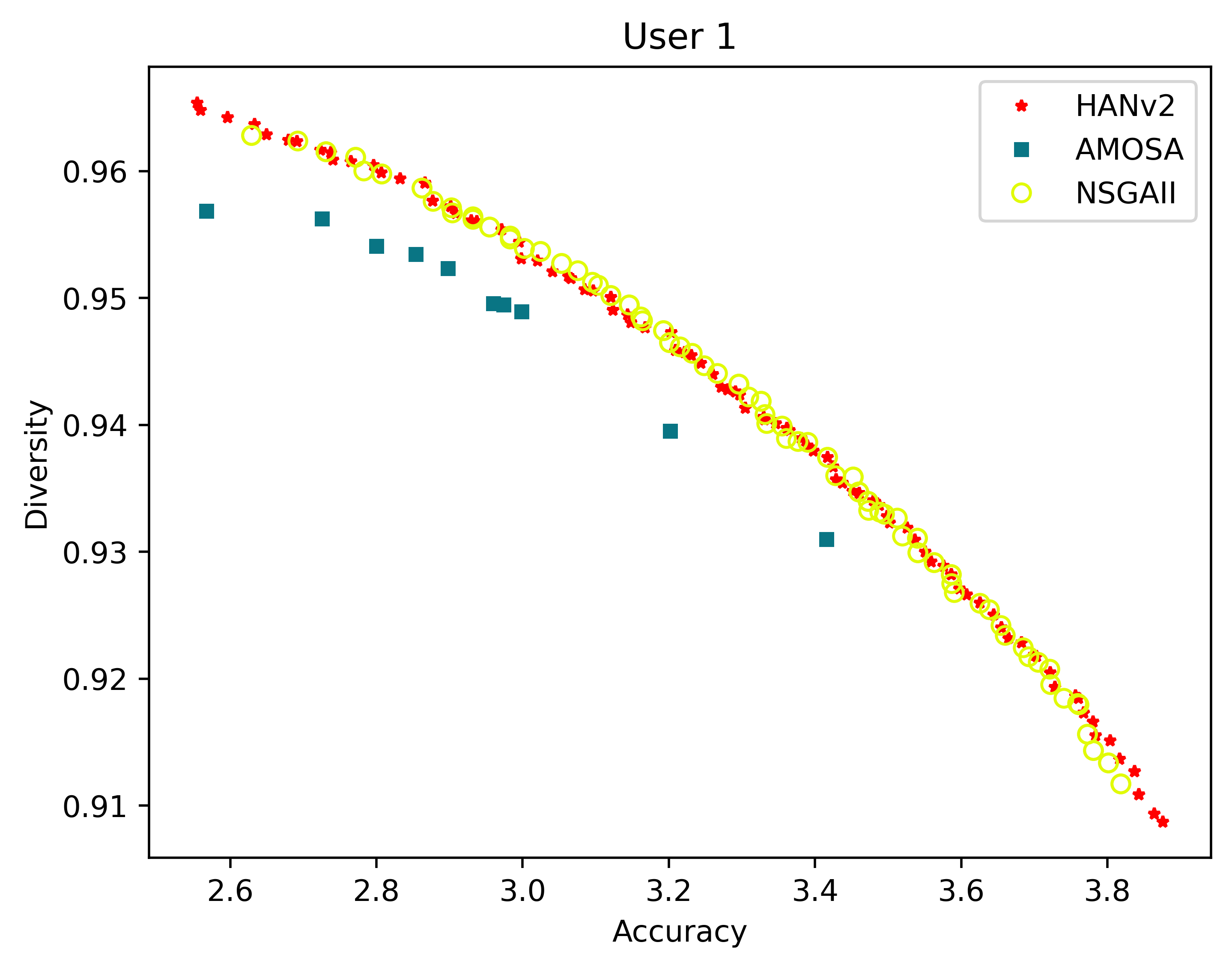}}\quad
        {\includegraphics[width= 4cm]{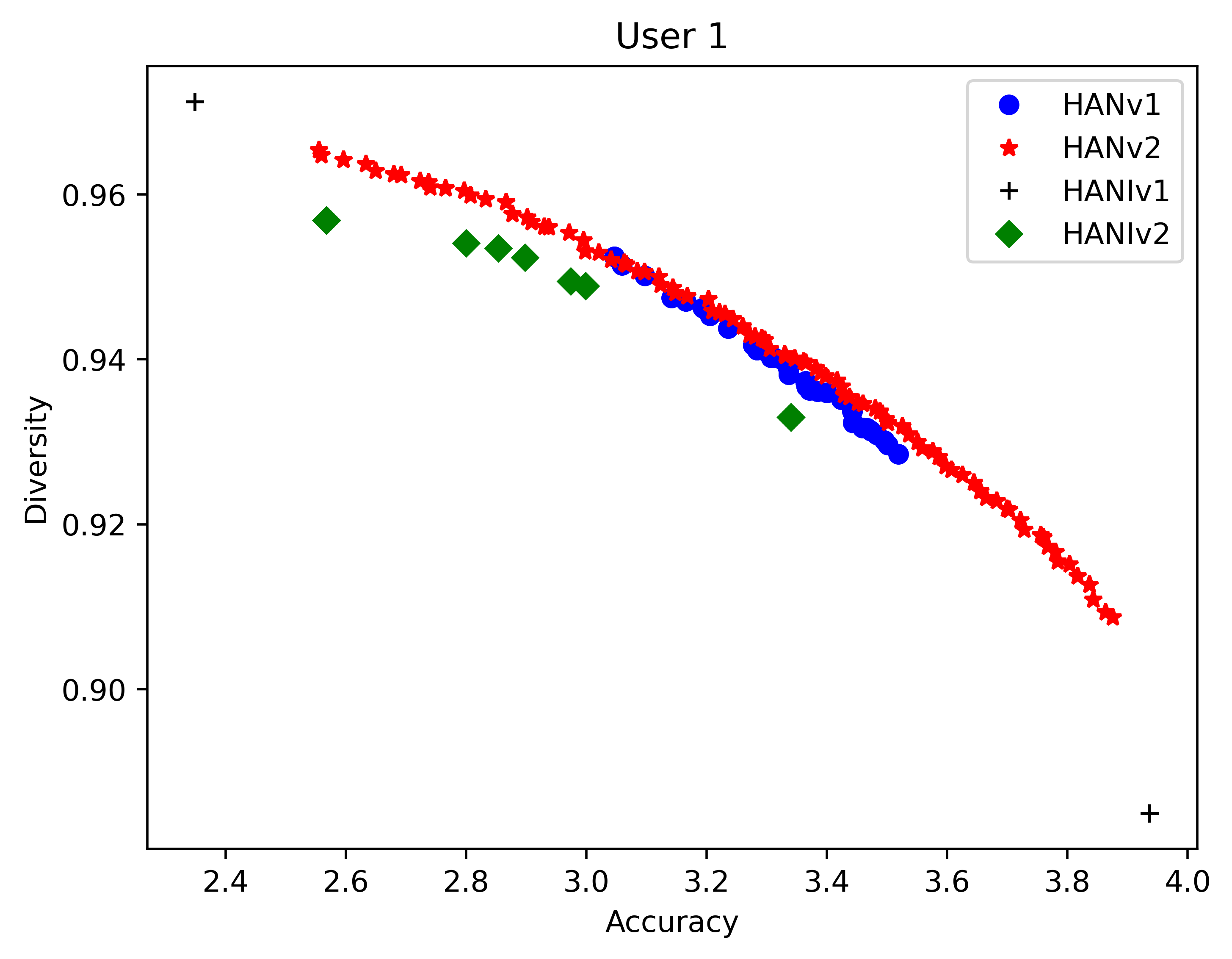}}
        \caption{
        Pareto frontier comparisons for user 1 (ModCloth).}
         \label{fig:paracom:modu1}
    \end{figure*}

\begin{figure*}[!h]
        {\includegraphics[width=4cm]{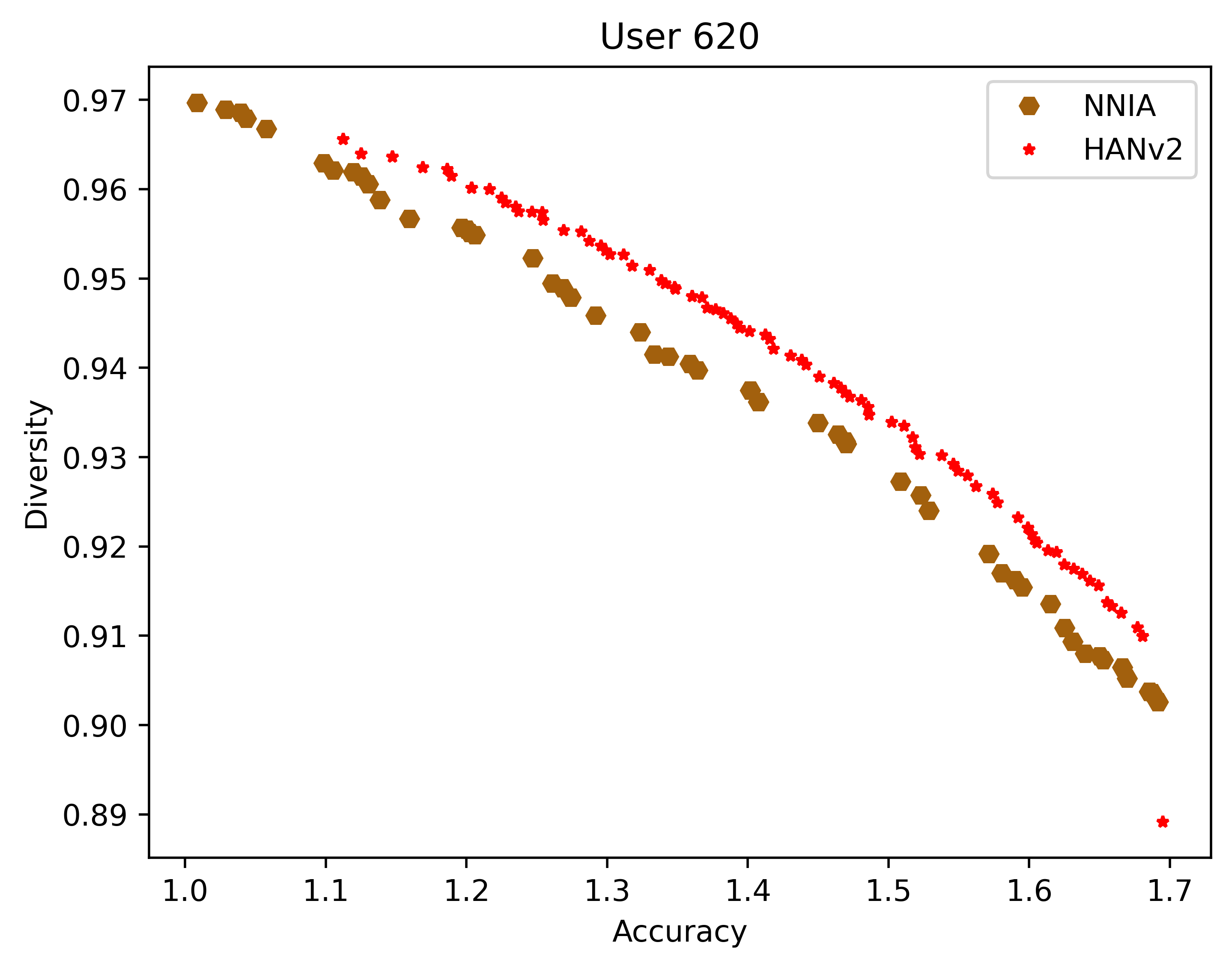}}\quad
        {\includegraphics[width= 4cm]{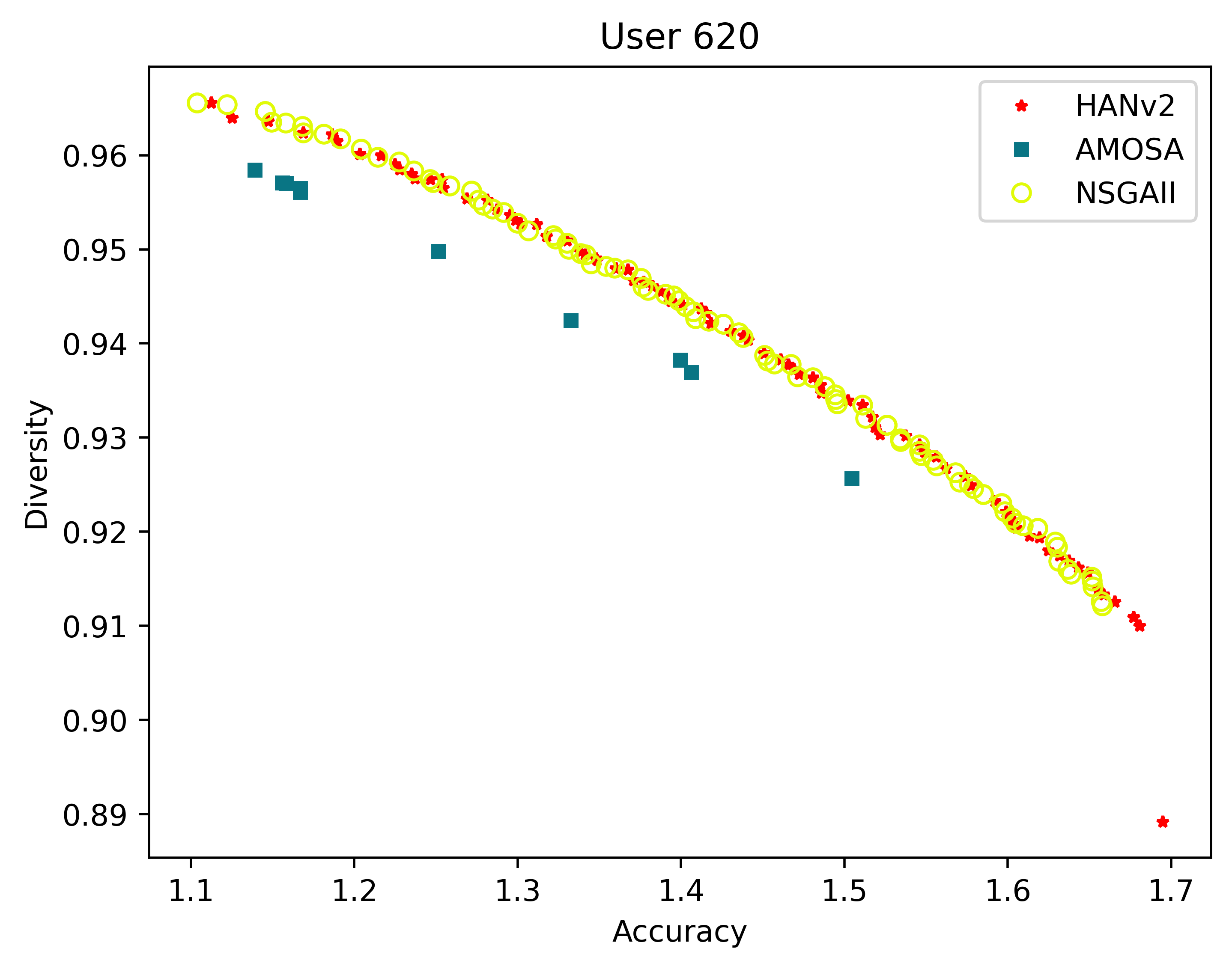}}\quad
        {\includegraphics[width= 4cm]{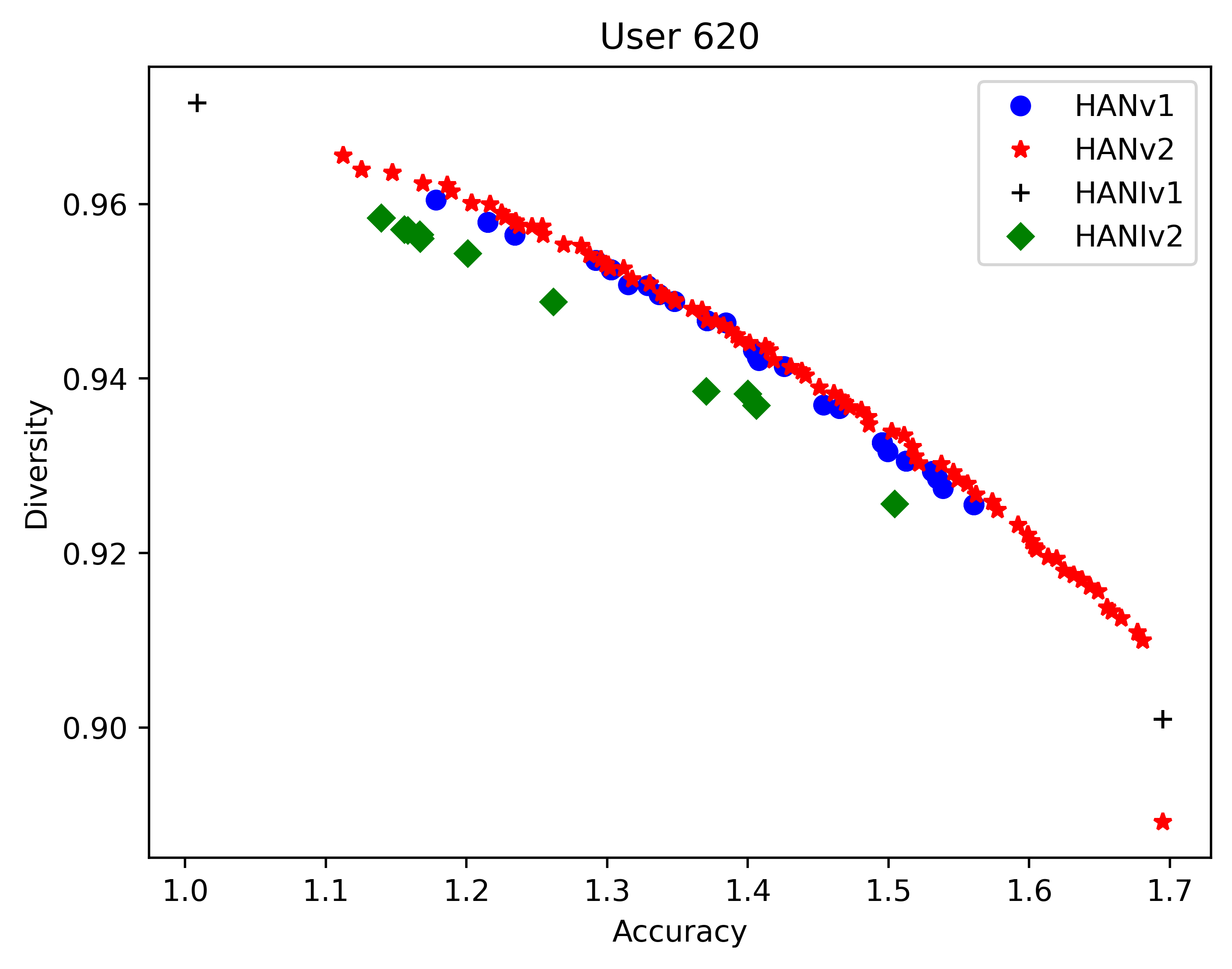}}
        \caption{
        Pareto frontier comparisons for user 620 (ModCloth).}
         \label{fig:paracom:modu620}
    \end{figure*}

Regarding the ModCloth dataset, Tables \ref{tab:smmid:Mod}, \ref{tab:dmsns:Mod}, and \ref{tab:clo:Mod}, along with Figures \ref{fig:paracom:modu1} and \ref{fig:paracom:modu620}, also illustrate the superiority of HANv2 over the other multi-objective algorithms.

 \begin{table*}[!h]
\caption{\label{tab:smmid:MovieLens} SM and MID on MovieLens.}
   \scalebox{0.55}{
    \begin{tabular*}{\linewidth}{@{\extracolsep{\fill}}cccccccccccccccccccccccccccccc@{}}
\cmidrule(lr){1-21}
UserID  &\multicolumn{2}{c}{NNIA}&& \multicolumn{2}{c}{NSGAII}&& \multicolumn{2}{c}{AMOSA}&& \multicolumn{2}{c}{HANv1}&& \multicolumn{2}{c}{HANv2}&& \multicolumn{2}{c}{HANIv1}&& \multicolumn{2}{c}{HANIv2}\\
\cmidrule(lr){2-3}\cmidrule(lr){5-6}\cmidrule(lr){8-9}\cmidrule(lr){11-12}\cmidrule(lr){14-15}\cmidrule(lr){17-18}\cmidrule(lr){20-21}
          &SM & MID && SM & MID &&SM  & MID && SM & MID && SM & MID &&SM  & MID && SM & MID &&\\
\cmidrule(lr){1-21}
1  &0.176 & 0.7646 && 0.048 & 0.7241 && 0.1742 & 0.7278 && 0.0882 & 0.7383 && 0.0507 & 0.7292 && 2.3624 & 0.9963 && 0.3226 & 0.7704  \\ 
1000  &0.2117 & 0.7898 && 0.0686 & 0.7431 &&0.1831 & 0.7969 &&0.1147 & 0.7642&&0.0775 & 0.7276 && 6.9549 & 1.0 &&0.3458 & 0.7568 \\
3411  &4.8152 & 0.986 &&0.2303 & 0.7274 && 0.4797 & 0.7423&& 0.5343 & 0.695 && 0.2477 & 0.7041&& 12.3586 & 0.9897&& 19.1856 & 1.0 \\ 
3415 &5.1145 & 0.9961 &&0.131 & 0.7237 &&0.3856 & 0.703 &&0.2398 & 0.7572 &&0.1537 & 0.7397 && 7.4074 & 0.9903 && 0.4338 & 0.7501 \\
3420 & 0.4353 & 0.7638 &&0.1437 & 0.7237&&0.3027 & 0.6748 &&0.2324 & 0.7446&&0.1576 & 0.6593&&2.9308 & 0.927&&0.304 & 0.6984\\  
\cmidrule(lr){1-21}
\end{tabular*}}
\end{table*}

 \begin{table*}[!h]
\caption{\label{tab:dmsns:MovieLens} DM and SNS on MovieLens.}
   \scalebox{0.62}{
    \begin{tabular*}{\linewidth}{@{\extracolsep{\fill}}cccccccccccccccccccccccccccccc@{}}
\cmidrule(lr){1-21}
UserID  &\multicolumn{2}{c}{NNIA}&& \multicolumn{2}{c}{NSGAII}&& \multicolumn{2}{c}{AMOSA}&& \multicolumn{2}{c}{HANv1}&& \multicolumn{2}{c}{HANv2}&& \multicolumn{2}{c}{HANIv1}&& \multicolumn{2}{c}{HANIv2}\\
\cmidrule(lr){2-3}\cmidrule(lr){5-6}\cmidrule(lr){8-9}\cmidrule(lr){11-12}\cmidrule(lr){14-15}\cmidrule(lr){17-18}\cmidrule(lr){20-21}
           &DM & SNS &&DM & SNS && DM & SNS && DM & SNS && DM & SNS && DM & SNS && DM & SNS &&\\
\cmidrule(lr){1-21}
1  & 5.27 &9.69 && 4.16 & 10.09 && 4.35 & 10.94 && 3.0 & 9.51 && 4.44 & 10.18 && 4.72 & 10.9 && 4.19 & 10.35  \\ 
1000  &7.4 & 18.11&&6.58 & 18.09&&6.22& 18.27&&3.9 & 18.32&&7.13 & 18.1&&6.95 & 25.0&&6.22 & 18.86\\
3411  &24.07 & 76.44&&20.26 & 68.12&&19.19 & 65.62&&18.17 & 67.61&&20.8 & 68.42&&24.72& 75.7&&19.19 & 94.6\\
3415 &15.34 & 36.17&&12.18 & 36.45&&13.88 & 37.08&&7.67 & 36.75&&13.98 & 36.17&&14.81 & 40.16&&13.88 & 36.59\\
3420 & 14.36 & 39.82&&12.64 & 38.94&&12.71 & 38.11&&7.67 & 40.02&&14.49 & 38.89&&14.65 & 43.67&&12.46 & 39.04\\  
\cmidrule(lr){1-21}
\end{tabular*}}
\end{table*}

 \begin{table*}[!h]
\caption{\label{tab:clo:MovieLens} CLO and Rank on MovieLens.}
   \scalebox{0.57}{
    \begin{tabular*}{\linewidth}{@{\extracolsep{\fill}}cccccccccccccccccccccccccccccc@{}}
\cmidrule(lr){1-21}
UserID  &\multicolumn{2}{c}{NNIA}&& \multicolumn{2}{c}{NSGAII}&& \multicolumn{2}{c}{AMOSA}&& \multicolumn{2}{c}{HANv1}&& \multicolumn{2}{c}{HANv2}&& \multicolumn{2}{c}{HANIv1}&& \multicolumn{2}{c}{HANIv2}\\
\cmidrule(lr){2-3}\cmidrule(lr){5-6}\cmidrule(lr){8-9}\cmidrule(lr){11-12}\cmidrule(lr){14-15}\cmidrule(lr){17-18}\cmidrule(lr){20-21}
           &CLO & Rank &&CLO & Rank && CLO & Rank && CLO & Rank && CLO & Rank && CLO & Rank && CLO & Rank &&\\
\cmidrule(lr){1-21}
1  & 0.9280 & 4 && 0.9428 & 2 && 0.9322 & 3 && 0.8914 &5 && 0.9541 & {\color{blue}\textbf{1}} && 0.0873 & 7 &&  0.8709 & 6 \\ 
1000  &0.8770& 4 && 0.8788 & 3 && 0.8761 & 5 && 0.8539& 6 && 0.8806&  {\color{blue}\textbf{1}} && 0.1415 & 7  && 0.8790 & 2   \\ 
3411  &0.7320 & 5 &&0.8525&2  && 0.8369& 4 && 0.8420&3  && 0.8552&{\color{blue}\textbf{1}}  && 0.3626& 6  && 0.1530 & 7   \\
3415 & 0.3323& 6 &&0.9313& 4 && 0.9433& 2 && 0.8703& 5 && 0.9461& {\color{blue}\textbf{1}} && 0.1216& 7 && 0.9340 & 3  \\
3420 & 0.8864 & 5 && 0.9425 & 2 &&0.9163 & 4 &&0.8848 & 6 && 0.9532 & {\color{blue}\textbf{1}} &&0.1125& 7 && 0.9191 & 3   \\
\cmidrule(lr){1-21}
\end{tabular*}}
\end{table*}

 \begin{table*}[!h]
\caption{\label{tab:smmid:Mod} SM and MID on ModCloth.}
   \scalebox{0.55}{
    \begin{tabular*}{\linewidth}{@{\extracolsep{\fill}}cccccccccccccccccccccccccccccc@{}}
\cmidrule(lr){1-21}
UserID  &\multicolumn{2}{c}{NNIA}&& \multicolumn{2}{c}{NSGAII}&& \multicolumn{2}{c}{AMOSA}&& \multicolumn{2}{c}{HANv1}&& \multicolumn{2}{c}{HANv2}&& \multicolumn{2}{c}{HANIv1}&& \multicolumn{2}{c}{HANIv2}\\
\cmidrule(lr){2-3}\cmidrule(lr){5-6}\cmidrule(lr){8-9}\cmidrule(lr){11-12}\cmidrule(lr){14-15}\cmidrule(lr){17-18}\cmidrule(lr){20-21}
          &SM & MID && SM & MID && SM & MID && SM & MID && SM & MID && SM & MID && SM & MID &&\\
\cmidrule(lr){1-21}
1   &0.7604 & 0.9967 && 0.0168 & 0.7214 && 0.0943 & 0.7355 && 0.0153 & 0.7906 && 0.0147 & 0.7283 && 1.59 & 1.0 && 0.1288 & 0.7253 \\
620  & 0.014 & 0.7989 && 0.0065 & 0.7444 & &0.0408 & 0.8564 & &0.0167 & 0.7705 && 0.0074 & 0.654 & &0.6899 & 1.0 &&0.0367 & 0.8566 \\   
624 & 0.0057 & 0.7412 &&0.0012 & 0.7312 && 0.0333 & 1.0 && 0.0016 & 0.747 && 0.0013 & 0.7016 && 0.0532 & 0.9993 && 0.0323 & 1.0 \\ 
629 & 2.0449 & 0.9861 && 0.0397 & 0.6645 && 0.1269 & 0.6522 && 0.0652 & 0.7741 && 0.0436 & 0.6435 &&1.951 & 0.9962 && 0.1396 & 0.764 \\
1000  &0.0483 & 0.9819 && 0.0009 & 0.6968 && 0.0125 & 0.8576 && 0.0012 & 0.7625 && 0.0013 & 0.6162 & &0.0459 & 0.9918 & &0.0445 & 1.0 \\
\cmidrule(lr){1-21}
\end{tabular*}}
\end{table*}

  \begin{table*}[!h]
\caption{\label{tab:dmsns:Mod} DM and SNS on ModCloth.}
   \scalebox{0.7}{
    \begin{tabular*}{\linewidth}{@{\extracolsep{\fill}}cccccccccccccccccccccccccccccc@{}}
\cmidrule(lr){1-21}
UserID  &\multicolumn{2}{c}{NNIA}&& \multicolumn{2}{c}{NSGAII}&& \multicolumn{2}{c}{AMOSA}&& \multicolumn{2}{c}{HANv1}&& \multicolumn{2}{c}{HANv2}&& \multicolumn{2}{c}{HANIv1}&& \multicolumn{2}{c}{HANIv2}\\
\cmidrule(lr){2-3}\cmidrule(lr){5-6}\cmidrule(lr){8-9}\cmidrule(lr){11-12}\cmidrule(lr){14-15}\cmidrule(lr){17-18}\cmidrule(lr){20-21}
          &DM & SNS &&DM & SN && DM & SN && DM & SN && DM & SN && DM & SN && DM & SN &&\\
\cmidrule(lr){1-21}
1   &1.52 & 3.22&& 1.19 & 2.76&& 0.85 & 2.49&& 0.47 & 2.72&& 1.32 & 2.73&& 1.59 & 3.4&&  0.77 & 2.54\\
620  &0.69 & 0.9& &0.56 & 0.98& &0.37 & 0.78& &0.38 & 0.94& &0.59 & 1.07& & 0.69 & 1.0& &0.37 & 0.77\\
624 & 0.11 & 0.22&&0.09 & 0.23&&0.03 & 0.07&&0.05 & 0.21&&0.09 & 0.26&&0.11 & 0.09&&0.03 & 0.06\\
629 & 4.09 & 5.31&&3.12 & 5.53&&2.03 & 5.4&&2.09 & 5.35&&3.52 & 5.38&&3.9 & 5.38&&2.37 & 4.3\\
1000  &0.1 & 0.05 &&0.07 & 0.26 &&0.05 & 0.11 &&0.04 & 0.2 &&0.1 & 0.35 &&0.09 & 0.07 &&0.04 & 0.07\\
\cmidrule(lr){1-21}
\end{tabular*}}
\end{table*}

 \begin{table*}[!h]
\caption{\label{tab:clo:Mod} CLO and Rank on ModCloth.}
   \scalebox{0.57}{
    \begin{tabular*}{\linewidth}{@{\extracolsep{\fill}}cccccccccccccccccccccccccccccc@{}}
\cmidrule(lr){1-21}
UserID  &\multicolumn{2}{c}{NNIA}&& \multicolumn{2}{c}{NSGAII}&& \multicolumn{2}{c}{AMOSA}&& \multicolumn{2}{c}{HANv1}&& \multicolumn{2}{c}{HANv2}&& \multicolumn{2}{c}{HANIv1}&& \multicolumn{2}{c}{HANIv2}\\
\cmidrule(lr){2-3}\cmidrule(lr){5-6}\cmidrule(lr){8-9}\cmidrule(lr){11-12}\cmidrule(lr){14-15}\cmidrule(lr){17-18}\cmidrule(lr){20-21}
           &CLO & Rank &&CLO & Rank && CLO & Rank && CLO & Rank && CLO & Rank && CLO & Rank && CLO & Rank &&\\
\cmidrule(lr){1-21}
1   &0.5440&6 &&0.8930& 2 && 0.8266& 3 && 0.8180& 5 && 0.9008& {\color{blue}\textbf{1}} && 0.1994&7 && 0.8142 &4\\
620  & 0.9264& 3 &&0.9401& 2 &&  0.8395& 5 &&  0.8842& 4 && 0.9651& {\color{blue}\textbf{1}}  && 0.1307&7  && 0.8394 &6   \\
624 & 0.8805& 3&& 0.9125& 2&&  0.3067& 6&& 0.8156& 4&& 0.9462&{\color{blue}\textbf{1}} && 0.2002&7 && 0.3155 &6\\
629 & 0.1717& 7 &&0.9223& 2 && 0.8382& 4 && 0.8445& 3 && 0.9510& {\color{blue}\textbf{1}} && 0.1748& 6 &&  0.8216&5   \\
1000  &0.1579&5 && 0.7874&2&& 0.4704&4 && 0.6595& 3 && 0.9942&  {\color{blue}\textbf{1}} &&  0.1505& 6 &&  0.0718 & 7 \\ 
\cmidrule(lr){1-21}
\end{tabular*}}
\end{table*}

\newpage
\section{Discussion, conclusion, and future work}
\label{sec:conclusion}

\subsection{Discussion}
This paper presented four novel hybrid multi-objective algorithms (HANv1, HANv2, HANIv1, and HANIv2) for recommendation systems, effectively combining the strengths of established algorithms such as AMOSA, NSGA-II, and NNIA. The experimental results from two datasets with varying characteristics, MovieLens and ModCloth, revealed several important findings.
\begin{itemize}
  \item Algorithm Performance:\\
 $-$ Robustness Across Datasets: HANv2 consistently outperformed the other algorithms across multiple evaluation metrics, particularly in generating well-balanced Pareto frontiers. While ICF approach demonstrated superior accuracy, the hybrid algorithms, especially HANv2, successfully achieved a more favorable balance between accuracy and diversity.
 \\
$-$ Adaptation to Data Sparsity: The algorithms exhibited different strengths depending on the density of the dataset. Notably, HANv2 showed remarkable robustness in the sparse ModCloth dataset, suggesting that it can effectively handle scenarios where user-item interactions are limited.

 \item Trade-off Management:
\\
$-$ The proposed hybrid approaches adeptly addressed the inherent accuracy-diversity dilemma. HANv2  excelled in maintaining high accuracy while simultaneously enhancing diversity in recommendations. This dual focus highlights the potential of multi-objective algorithms in real-world applications where user satisfaction often relies on a diverse set of options.
\\$-$ The three-stage framework implemented for generating and selecting optimal recommendation lists proved effective. By leveraging both recommendation quality metrics and Pareto frontier quality measures, the framework allowed for a comprehensive assessment of the recommender system.

\item Pareto Frontier Quality:
\\
$-$ The hybrid algorithms, particularly HANv2, generated more uniform and well-distributed Pareto frontiers compared to baseline approaches. This improved frontier quality not only reflects better trade-off management but also indicates the algorithms' capability to provide users with diverse recommendations that still adhere to their preferences.
\\$-$ The novel method for selecting optimal solutions from the Pareto set demonstrated efficacy in identifying balanced recommendations.
\end{itemize}

\subsection{Conclusion}
In conclusion, this work contributes  to the field of multi-objective recommendation systems through the following key outcomes:

\textbf{Novel Hybrid Algorithms}: The introduction of hybrid algorithms that successfully integrate the strengths of existing methods presents a valuable advancement in recommendation system design.

\textbf{Comprehensive Evaluation Framework:} The study establishes an evaluation framework that combines traditional recommendation metrics with measures of Pareto frontier quality, providing a more subtle understanding of algorithm performance.

\textbf{Empirical Evidence:} The findings provide empirical support for the effectiveness of hybrid approaches in managing the accuracy-diversity trade-off, demonstrating their potential for practical implementation in diverse recommendation scenarios.

\subsection{Future Work}
Building on the insights gained from this study, several directions for future research are suggested:
\\
$-$ \textbf{Multi-User Scalability:}  Extending the algorithms to handle multiple users simultaneously in a single run would enhance their practicality in real-world applications, where recommendations often need to provide to diverse user groups. 
\\
$-$ \textbf{Scalability for Larger Datasets:} Investigating the scalability of the proposed approaches for larger datasets is crucial, particularly given the increasing volume of data generated in various domains.
\\
$-$  \textbf{Additional Objectives:} Incorporating further objectives beyond accuracy and diversity, such as novelty or serendipity, could yield more holistic recommendation systems.
\\
$-$  \textbf{Adaptive Parameter Tuning}:  Developing mechanisms for adaptive parameter tuning could improve algorithm performance by dynamically adjusting to the characteristics of the dataset.
\\
$-$  \textbf{Domain-Specific Applications:}  Exploring the application of these algorithms in specific domains, such as news recommendation or e-commerce, would provide valuable insights into their versatility and effectiveness in varied contexts.

%
%


\bibliography{sn-bibliography}

\end{document}